\documentclass[10pt,journal,compsoc]{IEEEtran}
\usepackage{graphicx}%
\usepackage{multirow}%
\usepackage{amsmath,amssymb,amsfonts}%
\usepackage{amsthm}%
\DeclareMathOperator*{\argmax}{arg\,max}
\usepackage{mathrsfs}%
\usepackage{xcolor}%
\usepackage{textcomp}%
\usepackage{manyfoot}%
\usepackage{booktabs}%
\usepackage{algorithm}%
\usepackage{algorithmicx}%
\usepackage{algpseudocode}%
\usepackage{listings}%
\usepackage{subfig}
\usepackage{array}
\usepackage{hyperref}

\newcommand{\highlight}[1]{\textcolor{black}{#1}}

\ifCLASSOPTIONcompsoc
  \usepackage[nocompress]{cite}
\else
  \usepackage{cite}
\fi

\ifCLASSINFOpdf
\else
\fi

\begin{document}

\title{Could Micro-Expressions be Quantified? Electromyography Gives Affirmative Evidence}

\author{
        Jingting~Li,~\IEEEmembership{Member,~IEEE,}
        Shaoyuan~Lu,
        Yan~Wang,
        Zizhao~Dong,
        Su-Jing Wang\IEEEauthorrefmark{1},~\IEEEmembership{Senior Member,~IEEE,}
        and~Xiaolan~Fu,~\IEEEmembership{Member,~IEEE} 
\IEEEcompsocitemizethanks{
\IEEEcompsocthanksitem \IEEEauthorrefmark{1} Corresponding author.
\IEEEcompsocthanksitem This work is supported, in~part, by grants from the National Natural Science Foundation of China (62276252, 62106256), and in part, by~a grant from the Youth Innovation Promotion Association CAS. 
\IEEEcompsocthanksitem  J.T Li, S.Y Lu, Y Wang, Z.Z Dong and S.J Wang are with the Key Laboratory of Behavior Sciences, Institute of Psychology, Chinese Academy of Sciences, Beijing, 100101, China, and also with the Department of Psychology, University of the Chinese Academy of Sciences, Beijing, 100049, China.\protect\\
E-mail: \{lijt, lusy, wangyan1, dongzz, wangsujing\}@psych.ac.cn
\IEEEcompsocthanksitem  X.L Fu is with the State Key Laboratory of Brain and Cognitive Science, Institute of Psychology, Chinese Academy of Sciences, Beijing, 100101, China, and also with the Department of Psychology, University of the Chinese Academy of Sciences, Beijing, 100049, China.
\protect\\
E-mail: fuxl@psych.ac.cn
}}

\markboth{Journal of \LaTeX\ Class Files,~Vol.~14, No.~8, August~2015}%
{Li \MakeLowercase{\textit{et al.}}: Could Micro-Expressions be Quantified? Electromyography Gives Affirmative Evidence}

\IEEEtitleabstractindextext{%
\begin{abstract}
Micro-expressions (MEs) are brief, subtle facial expressions that reveal concealed emotions, offering key behavioral cues for social interaction. Characterized by short duration, low intensity, and spontaneity, MEs have been mostly studied through subjective coding, lacking objective, quantitative indicators. This paper explores ME characteristics using facial electromyography (EMG), analyzing data from 147 macro-expressions (MaEs) and 233 MEs collected from 35 participants. First, regarding external characteristics, we demonstrate that MEs are short in duration and low in intensity. 
Precisely, we proposed an EMG-based indicator, the percentage of maximum voluntary contraction (MVC\%), to measure ME intensity. Moreover, we provided precise interval estimations of ME intensity and duration, with MVC\% ranging from 7\% to 9.2\% and the duration ranging from 307 ms to 327 ms. This research facilitates fine-grained ME quantification.
Second, regarding the internal characteristics, we confirm that MEs are less controllable and consciously recognized compared to MaEs, as shown by participants’ responses and self-reports. This study provides a theoretical basis for research on ME mechanisms and real-life applications. 
\highlight{Third, building on our previous work, we present CASMEMG, the first public ME database including EMG signals, providing a robust foundation for studying micro-expression mechanisms and movement dynamics through physiological signals.}
\end{abstract}

}

\maketitle

\IEEEdisplaynontitleabstractindextext

%
\IEEEpeerreviewmaketitle

\IEEEraisesectionheading{\section{Introduction}\label{sec:introduction}}



\IEEEPARstart{R}{ecognizing} emotions is essential for effective engagement in social interactions~\cite{darwin1872ausdruck}. 
Among the various ways emotions are expressed, facial expressions are widely regarded as the most intuitive way to recognize emotions and interpret intentions~\cite{stewart2009presidential}. 
When individuals intend to suppress or conceal their emotions, micro-expressions (MEs) may leak out on face~\cite{ekman2006flawed,ekman2009lie,yan2013fast}. Specifically, 
\highlight{MEs and macro-expressions (MaEs) are fundamentally distinct in their mechanisms and external appearances. MEs are brief, involuntary facial movements lasting less than 500 milliseconds, often subtle and easily missed, and typically occur in localized facial areas with potential asymmetry~\cite{liu2021micro,hong2019characterizing}. In contrast, MaEs are longer, more intense, consciously observable, involve symmetric movements, and engage a broader range of facial muscles.}
\par
As non-contact and imperceptible cues, ME can reflect an individual's true emotions and have potential applications in fields such as healthcare and public safety.
However, manually analyzing MEs can be time-consuming and labor-intensive, and the results may be influenced by subjective factors. On the other hand, computer vision-based ME analysis, which identifies true emotions by extracting motion characteristics of MEs, can be widely, conveniently, and effectively applied in various scenarios.
\par
In 2009, Polikovsky et al. made the first attempt to automatically recognize MEs using 3D-HOG features~\cite{polikovsky2009facial}. Later, with the release of databases such as SMIC~\cite{li2013spontaneous} and CASME series~\cite{yan2013casme,yan2014casme}, computer vision-based ME recognition gradually emerged. Then, with the continuous development of deep learning models, ME recognition algorithms have gradually evolved from initial hand-crafted-feature-based machine learning methods to a variety of deep learning algorithms. Meanwhile, since 2016, the release of long-video ME databases such as CAS(ME)$^2$~\cite{qu2017cas} and SAMM-LV~\cite{yap2020samm} has promoted research on ME spotting. As MEs can easily be confused with other types of facial movements or environmental noise, ME spotting is still in its infancy, and its performance is not yet satisfactory~\cite{li2022megc2022,davison2023megc2023}. 
\par
Although the analysis of MEs is rapidly developing, the three ME characteristics mentioned at the beginning still requires further clarification. Previous definitions based on ME samples predominantly relied on visual observation, a method relatively subjective. Specifically, the Facial Action Coding System (FACS)~\cite{ekman1971facial,ekman1978facial,ekman1982felt}, a well-established coding framework, offers effective methods for analyzing both action units (AUs) and emotions. However, the intensity coding in FACS relies solely on subjective ratings by coders~\cite{cohn2007observer}, which can lead to weak consistency. Consequently, the analysis of ME characteristics lacks objective and replicable quantification method . 

\par
For external characteristics of MEs, they can be delineated using objective physiological indicators. For instance, electromyography (EMG), a method known for its high accuracy in physiological signal acquisition~\cite{dimberg2002facial}, would be particularly effective in objectively measuring external aspects of MEs, such as their intensity and duration. Employing objective characterization methods can facilitate the rapid construction of large-scale ME databases, alleviating the small sample size issue prevalent in current intelligent analysis. Moreover, a clear understanding of the external features of MEs can provide essential prior knowledge for ME analysis based on deep learning.
\par
Simultaneously, the internal characteristics are also a crucial component for MEs. Precisely definition not only makes the ME coding more reliable but also lays a theoretical foundation for the identification of hidden, genuine emotions in practical applications based on MEs. While the previous study has established a correlation between self-awareness of facial expressions and external characteristics~\cite{qu2017you}, it remains to be confirmed if the spontaneity of MEs aligns with this conclusion.

\par

\begin{figure*}
    \centering
    \includegraphics[width=0.8\linewidth]{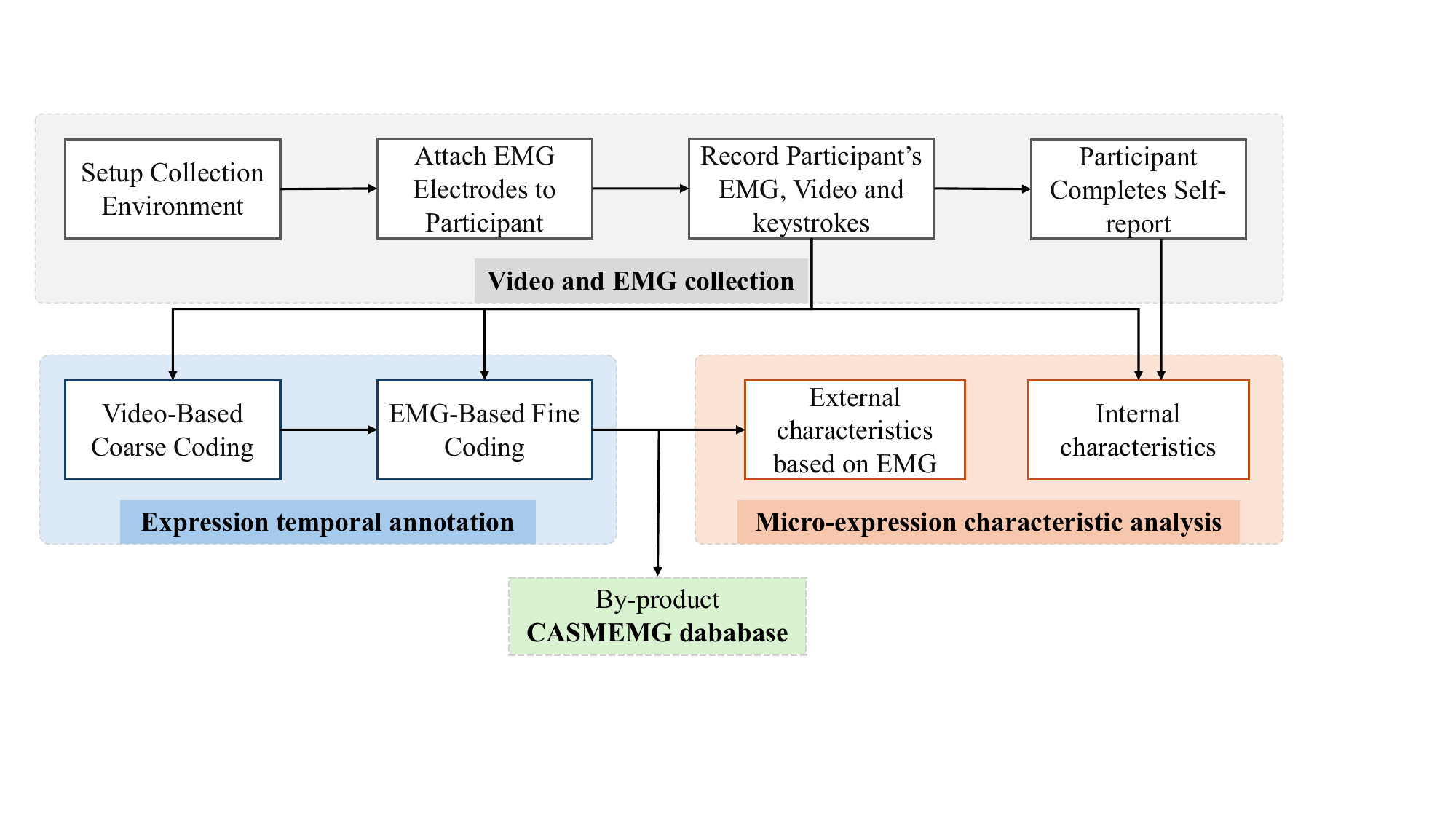}
    \caption{Overview of EMG-based ME characteristic quantification. Our research process is divided into three main parts: Video and EMG collection, Expression temporal annotation and ME characteristic analysis. During this process, a ME database containing EMG and video, named CASMEMG, was naturally developed.}
    \label{fig:pipeline}
\end{figure*}

\highlight{Overall, the motivation for this paper comes from two main aspects. First, the theoretical basis for MEs is largely hypothetical, with few studies quantifying their characteristics through physiological signals. Second, the undefined external characteristics of MEs make it difficult for deep learning models to learn and extract effective features.} To address these challenges, we propose a quantitative study of MEs using EMG and self-reports as direct indicators.
This paper is an extended version of an article we published in the Facial ME workshop at the ACM International Conference on Multimedia~\cite{lu2022EMG}. Fig.~\ref{fig:pipeline} illustrates the overview of this study. In particular, we estimated intervals for the duration and intensity of MEs. Additionally, we utilized a self-report method to investigate the spontaneity of MEs. Our findings support prior definitions of MEs and provide objective indicators for further analysis. \highlight{Besides, We organized all the collected data to build an EMG-based ME database called CASMEMG.}  The contributions of this article can be summarized as follows.
\begin{itemize}
    \item Firstly, we present objective and dependable findings from facial EMG-based ME studies, indicating that MEs are characterized by shorter duration and lower intensities. Notably, we introduce EMG-based measures as novel indicators of ME intensity. Additionally, we offer precise interval estimations for the duration and intensity of MEs, facilitating a refined quantification of external characteristics.
   
    \item Secondly, we conducted an empirical study to investigate the internal characteristics of MEs. We found that individuals exhibit less control and awareness over MEs compared with macro-expressions (MaEs). This research provides the theoretical basis for subsequent studies on ME mechanisms and applications in real-life scenarios.
    
    \item \highlight{Thirdly, we have also created a ME database: \textbf{CASMEMG}\footnote{The CASMEMG database is available for application by submitting the license agreement through \url{http://casme.psych.ac.cn/casme/e5}}, which features EMG signals as objective physiological indicators alongside video recordings. It contains 380 samples of MEs and MaEs in total. Keyframe timings, AUs, and emotion labels are all provided. Additionally, a baseline method and results for EMG-based expression interval detection are included. The release of CASMEMG database aims to enhance research into the underlying mechanisms of micro-expression movements.} 
\end{itemize}

\section{Background}

\subsection{Duration}

Duration is usually used as a criterion to distinguish MEs from MaEs. The different boundaries for ME duration have been given in various research and are systematically presented in the later sub-sections.

\textbf{Hypothetical Definition:}
In the early research of MEs, most definitions of MEs came from Ekman's research team. 
In particular, the earliest definition was put forward in 1969. Ekman et al. proposed that MEs were subtle and rapid facial movements, lasting less than 1/2~s~\cite{ekman1969nonverbal}. Conversely, MaEs were mostly between 1/2~s and 1~s. 
Subsequently, Ekman et al. proposed that when emotion was hidden or masked, the genuine emotion was expressed as a ME, a transient expression lasting between 1/25~s and 1/5~s~\cite{ekman1975unmasking}.
Among these two definitions, the duration of 1/5~s is more agreed upon in the later ME experiments and is considered the classical definition. 
Besides, Ekman et al. had proposed definitions of ME duration of 1/4~s~\cite{ekman1985telling,ekman2003darwin} and 1/3~s~\cite{ekman2005face} in other books, respectively. 
Nevertheless, the duration definitions of MEs listed above are hypothetical classifications that have rarely been directly verified by empirical research.

\textbf{Validation Definition:}
%
In addition to hypothetical definitions, researchers have attempted to explore the duration definition of MEs through empirical approaches.
The first experiment that demonstrated the existence of MEs was done by Porter et al.~\cite{porter2008reading}. The duration of MEs in this study matched Ekman's earlier definition (1/5~s -1/2~s)~\cite{ekman1975unmasking}. 
Besides the psychological approach, based on 3D gradient descriptors and FACS coding methods, Polikovsky et al. indicated that the duration of MEs is between 1/25~s and 1/3~s~\cite{polikovsky2009facial}.
Later, Yan et al. explored the distribution of leaked fast facial expressions and gave duration bounds for MEs with a lower limit of about 170~ms and an upper limit of about 500~ms. They also concluded that the duration of the onset phase, i.e., from the onset to the apex, could be a suitable measure for defining MEs with a lower limit of about 65~ms and an upper limit of about 260~ms~\cite{yan2013fast}. 
However, these precise definitions of ME duration are derived based on FACS. Thus, the conclusions obtained may be inconsistent due to the subjective coding process.

\textbf{Operational Definition:}
Due to the unknown nature of the process of ME production, researchers have also explored the operational definitions of MEs and related evidence from the recognition perspective. For instance, Ekman et al. in 1991 validated their proposed ME duration definition based on mannual ME recognition. Specifically, they used MEs of 1/25~s to explore the correlation between ME recognition rate and the accuracy of deception detection~\cite{ekman1991can}. 
In addition, Shen et al. found that 200~ms was a turning point for recognition rate in exploring the relationship between ME recognition and duration~\cite{shen2012effects}. When participants observed MEs in pictures with a presentation duration of less than 200~ms, the recognition rate increased with increasing presentation duration. In contrast, the recognition rate remained the same when the presentation duration of the 
expression pictures was 200~ms or more. In sum, the boundary point between the duration of MEs and MaEs was 200~ms.
\par
Yet, the operational definition established from the perspective of ME recognition is different from the definition of ME production. Only MEs occurring for a long enough duration can be self-perceived by the individual. The recognition of MEs by others and the self-perception of MEs are derived from different cognitive processing and thus will yield different results.

\textbf{Quantitative Definition:}
Although the definition of the duration of MEs has evolved from approximate estimation to manual frame-by-frame annotation in videos, the consistency of manual coding among coders is challenging.  Specifically, researchers have suggested that when the video is running at normal speed, coders are unaware of facial changes occurring within 1/5~s and can only recognize "something is happening" for 2/5~s but cannot identify the specific expression~\cite{haggard1966micromomentary}. 
Therefore, in manual annotation, coders need to observe the video repeatedly at a slow speed to detect the occurrence of MEs. In addition, due to the subtle and transient characteristics of MEs, there are differences in perception ability, subjective definition, and fluctuation of attention among coders when coding the onset and offset frames. These reasons could lead to unreliable and inconsistent manual annotation results.
\par

With the development of high-precision physiological signal detection equipment and artificial intelligence, the duration of MEs can be measured by more objective indicators besides manual coding. 
Recently, Kim et al. employed electroencephalography (EEG) to detect MEs. \highlight{However, even with high-resolution EEG, accurately estimating EEG power spectral features at low frequency bands such as theta and alpha requires a relatively long time window (1 s in the cited study). This time window reflects the characteristics of EEG signals over a period, not at a single moment. Consequently, it is not capable of providing information about the onset of MEs (less than 500 ms)~\cite{kim2022classification}.} In contrast, the temporal resolution of the EMG signal can reach milliseconds. 
In 2017, Perusquia-Hernandez et al. detected rapid and subtle smiles through a high-precision wearable EMG device~\cite{perusquia2017wearable}. 
Regarding the fact that these facial expressions are all expressed by facial muscle action, EMG is seen as an optimal choice for detecting ME and obtaining high precision duration.


\subsection{Intensity}

Although low intensity is considered one of the characteristics of MEs, there is no compelling empirical approach to demonstrate a significant intensity difference between the ME and MaE. Currently, the popular ME annotation method is to code AUs based on FACS proposed by Ekman et al.~\cite{ekman1978facial}. In particular, FACS itself is descriptive, and FACS-based coders code expressions based primarily on observations of facial muscle actions. 
FACS provides a subjective classification standard for AU intensity, i.e., a descriptive scale of five grades, including distinguishing between trace (A), slight (B), marked or pronounced (C), severe or extreme (D), and maximum (E) intensities. 
Although FACS specifies intensity coding for many action units, it is difficult to establish and maintain an acceptable level of reliability, and further research is needed~\cite{cohn2007observer}. Furthermore, the FACS-based annotation of facial expressions usually requires a certified expert. In all, this approach is subjective and difficult to replicate, thus adding obstacles to the coding process. 
\par
In addition, there has been a lack of attention to intensity measurements for ME. The variation in AU intensity is sometimes treated as a separate event.  In subsequent coding of facial expressions based on AU combinations, the intensity of AU is ignored and is not reflected in the coding of expressions.  Moreover, in the field of computer science, relatively even less attention has been paid to intensity measurements for ME. Due to the limited number of ME databases with coded facial movement intensity, it is challenging to analyze ME intensities further~\cite{wang2018facial}.
\par
Meantime, facial EMG can detect the changes in the electromagnetic field potential during muscle contraction. Particularly, the maximum voluntary contraction (MVC)~\cite{lehman1999importance} of facial muscles can be quantified by EMG, and thus the muscle action intensity can be expressed as a percentage of MVC, ranging from resting muscles (0$\%$) to the maximum possible muscle contraction (100$\%$). 

\subsection{Spontaneity}
In the current research, the spontaneity of MEs is also one of the attention-grabbing characteristics. Specifically, the occurrence of MEs is not only difficult for observers to detect but also equally challenging to be controlled by oneself~\cite{warren2009detecting}. Accordingly, MEs are often thought not to appear in fabricated emotional expressions and cannot be suppressed or masked, thus providing valuable clues to the genuine emotions~\cite{johnson1975communicative}. Furthermore, the spontaneity of MEs has become attractive for applications such as deception detection~\cite {ekman1991can,frank1997ability,slowe2007automatic,yap2014facial}.

\par
The spontaneity of MEs is reflected in uncontrollability and unawareness. It can be inferred from the facial feedback hypothesis that individuals can be aware of expressions and emotions only when sensory feedback from facial muscles is strong enough to reach the consciousness~\cite{tomkins1962affect}. 
In other words, the subjective experience of facial expressions is correlated with the duration and intensity~\cite{adelmann1989facial}. 
Later, research confirmed that intensity is a significant predictor of self-awareness, which means that higher-intensity facial expressions are more likely to be aware by individuals~\cite{qu2017you}. 
Therefore, self-awareness of facial muscle actions is more difficult for MEs that are short in duration and lower in intensity. 
However, previous research has mainly focused on the application that spontaneous MEs can serve as a leakage of genuine emotions. 
The research on the cognition process behind the spontaneous characteristic of MEs is lacking. 
Through real-time self-report and subjective experience, differences in the production process of ME and MaE could be further explored.

\section{Methods}
\subsection{EMG-based External Characteristic Measurement}

As mentioned earlier, EMG is effective in capturing and quantifying facial muscle movements through various signal characteristics. In this study, we used EMG to objectively quantify the external characteristics of MEs, specifically focusing on their short duration and low intensity. We first synchronously collected EMG signals and facial videos. Using manually coded facial expressions from the video, we identified corresponding EMG segments for further analysis. The following subsections detail our EMG acquisition hardware, electrode distribution, signal pre-processing, and the EMG indicators used to characterize the external features of MEs, thereby clarifying our EMG-based methodology.

\subsubsection{EMG Acquisition Hardware}
To meet our experimental requirements, we designed a custom EMG acquisition module. Subsequent paragraphs will provide an in-depth overview of design of this module, including basic configuration, sampling rate, data transmission, and synchronization with video.
\par
\begin{figure}[b]
    \centering
    \includegraphics[width = \linewidth]{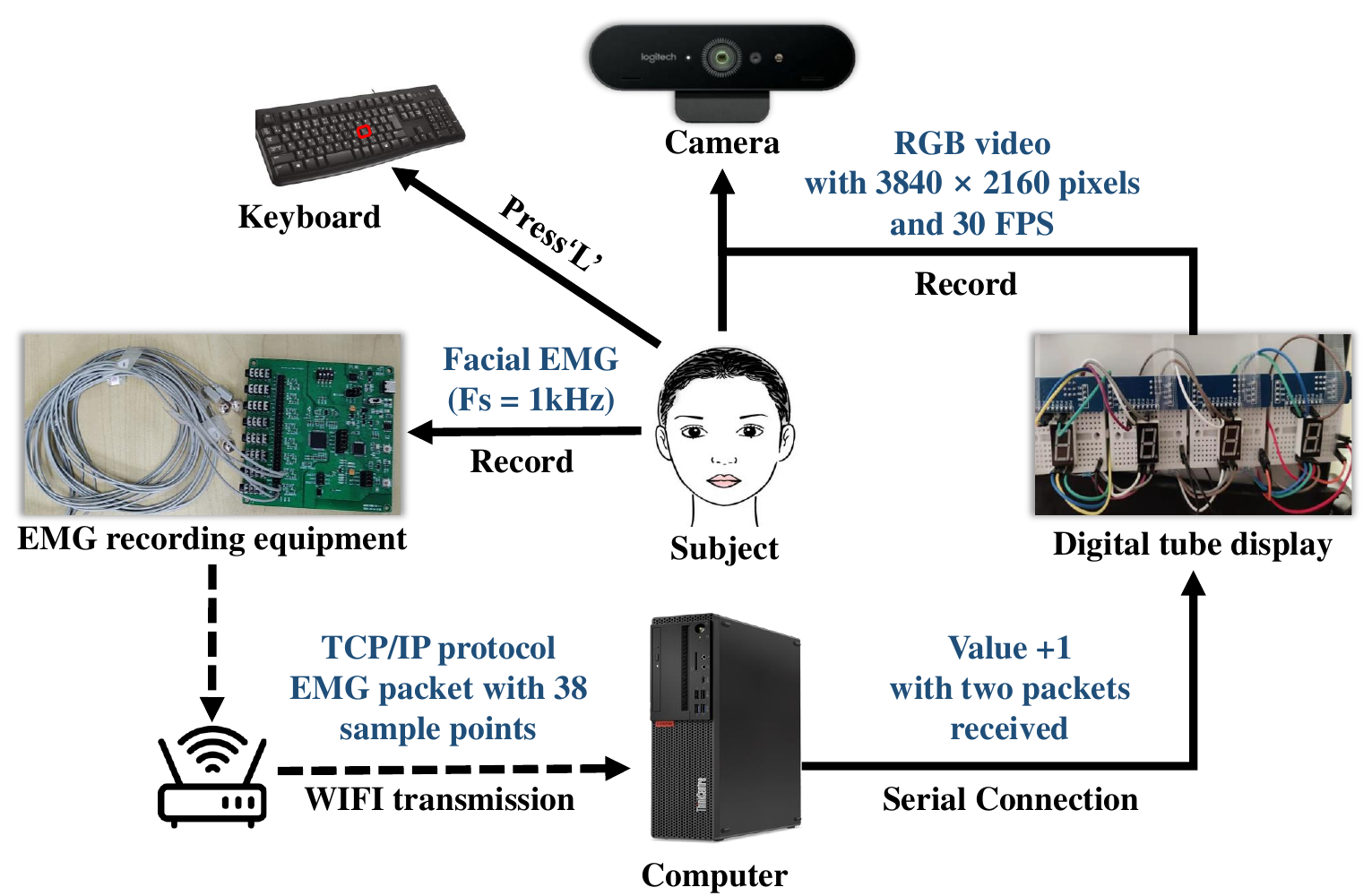}
    \caption{Equipment layout based on data collection and transmission }
    \label{fig:eqqLink}
\end{figure}

\begin{figure}[b]
  \centering
  \includegraphics[width=0.8\linewidth]{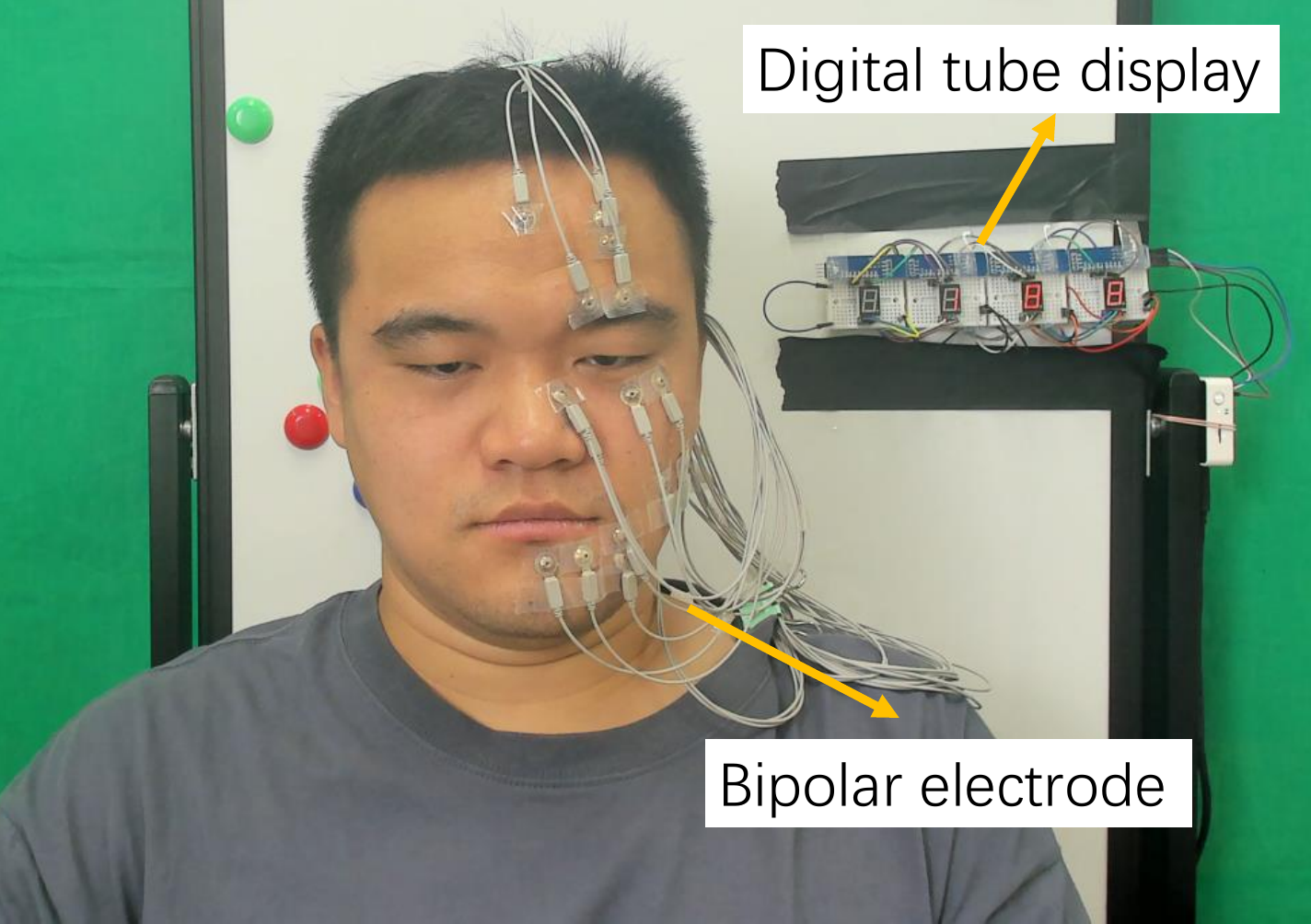}
  \caption{Data acquisition setup. Participants were recorded on video while watching emotional stimuli. Simultaneously, EMG signals were captured through electrodes attached to the participant's face and EMG module. Furthermore, the digital tube displays helped the coder synchronize the frames in the recorded video with the EMG signals. This example image is selected from recorded videos with the participant's consent.}
  \label{fig:scene}
\end{figure}
Regarding the EMG signal acquisition device, we needed to capture EMG signals from seven channels to ensure a comprehensive acquisition of facial muscle activities. Meantime, commercial EMG modules are typically limited to recording only two or three channels at a time. To fulfill our acquisition needs, we would have had to use three or four commercial EMG modules, which are both expensive and complex. Additionally, the underlying interfaces are not accessible to end-users. This makes it challenging to flexibly configure commercial EMG modules for experimental purposes.
Thus, to reduce costs and customize the device to our needs, we designed our own EMG module based on the Texas Instruments$^\circledR$ ADS1299 chip. In detail, it is an eight-channel, low-noise, 24-bit, simultaneous-sampling delta-sigma ($\Delta\Sigma$) analog-to-digital converters (ADCs) with a built-in programmable gain amplifier (PGA), internal reference, and an onboard oscillator~\cite{instruments2017ads1299}. 
Comparative experiments with commercial modules confirmed the reliability of our system, demonstrating its capability to meet our experimental needs~\cite{zhang2023perifacial}.
\par
Regarding the sampling rate, our EMG module allows sampling at 500 Hz, 1000 Hz, or 2000 Hz. For facial EMG signals, predominantly in the 20-500 Hz range~\cite{van2010facial}, we selected a 1000 Hz sampling rate. This rate meets the Nyquist–Shannon theorem~\cite{shannon1949communication}, ensuring accurate signal capture without aliasing. Sampling at 2000 Hz would introduce unnecessary noise, making 1000 Hz optimal for our experiments.

\par
Regarding the data transmission, Fig.~\ref{fig:eqqLink} shows the arrangement of acquisition devices and their transmission modes. To prevent interference from utility power, we used Wi-Fi to wirelessly transmit EMG data from our custom module to the computer. Specifically, the data was sent via a 2.4 GHz wireless router using the TCP/IP protocol. TCP, being connection-oriented and reliable, ensures consistent data transmission once a connection is established.

\par
For synchronizing the EMG signal with the video, we integrated a digital tube display into the video background, as illustrated in Fig.~\ref{fig:scene}. The digital tube displayed a changing number, captured alongside the participant’s facial muscle movements by a 4K camera, ensuring synchronization between the video frames and the EMG signal. Specifically, each time the PC received two EMG data packets, the value on the digital tube increased by one. The camera simultaneously recorded these changes and the participant’s facial movements, marking the continuous video with the digital tube’s updates. This allowed the digital tube changes in the video to correspond directly to the sequence of EMG signals received by the PC, enabling precise timing between the EMG and video signals.

\par

\subsubsection{Electrode Distribution}
\begin{figure}[t]
  \centering
  \includegraphics[width=0.8\linewidth]{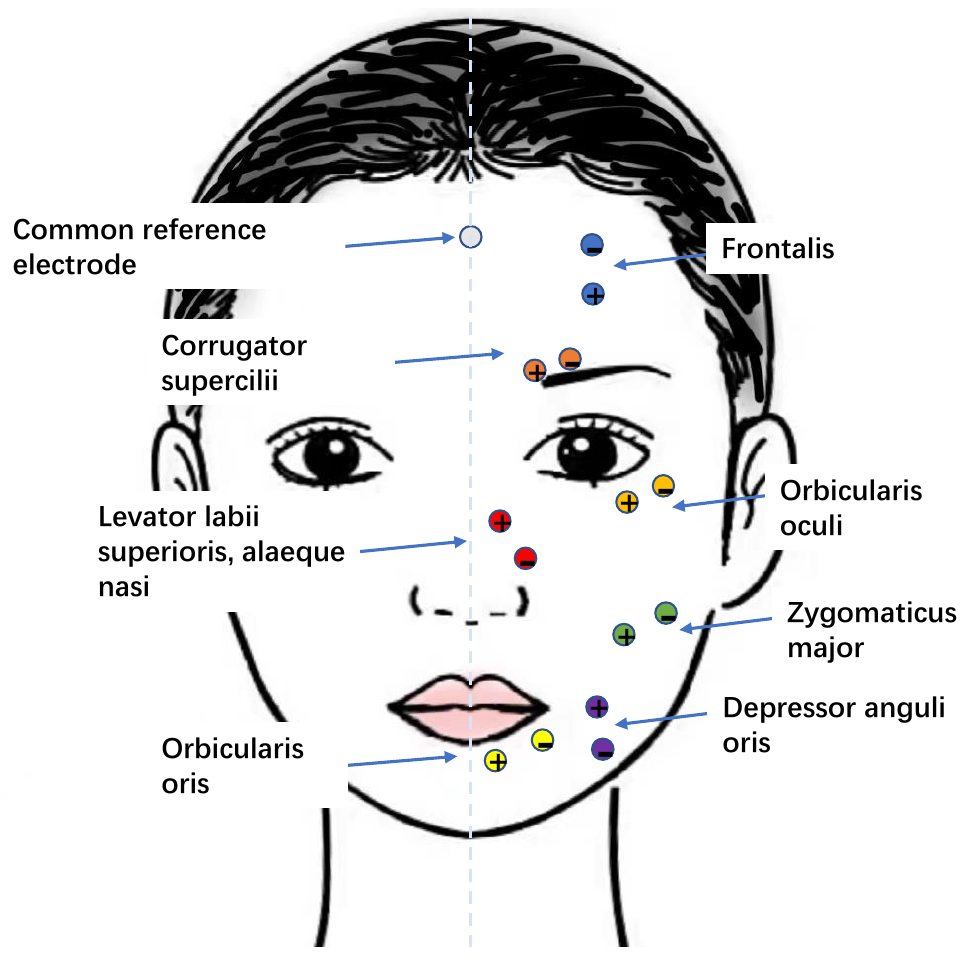}
  \caption{Electrode position distribution for facial EMG signal acquisition based on seven facial muscles }
  \label{fig:position}
\end{figure}

In our EMG setup, we acquire muscle action potentials by adhering \highlight{non-intrusive} electrode patches to the participant's face. \highlight{The captured signals are surface EMG, which is widely used in many fields related to muscle movement assessment, such as sports rehabilitation and exercise~\cite{campanini2020surface, mcmanus2020analysis,turker2013surface}.}
As shown in Fig.~\ref{fig:position}, bipolar electrodes were placed near the targeted muscle tissue. Each bipolar electrode consists of two units, one at the forward input and one at the backward input of the differential amplifier. This setup measures the potential difference between the two electrodes, effectively canceling out electrical signals from distant muscles. Consequently, compared to monopolar electrodes, the bipolar configuration is less prone to interference and crosstalk.

\par
Regarding the electrodes' distribution on the face, since we needed to balance ME coding and EMG signal acquisition, we chose the left half of the face to acquire EMG signals.  Previous studies have shown that the left facial region tends to express stronger emotions than the right~\cite{1984Asymmetry}. In particular, we selected seven muscles on the left side of the face as the central muscle regions for EMG signal acquisition, as shown in Fig.~\ref{fig:position}, namely frontalis (c1), corrugator supercilii (c2), orbicularis oculi(3), levator labii superioris alaeque nasi (c4), zygomaticus major (c5), orbicularis oris (c6), and depressor anguli oris (c7). \highlight{Specifically, as presented in Appendix~\ref{app:channel}, the seven facial muscles selected in this study largely encompass the range of AUs associated with the six basic emotions.}
The right side of the face was left without occlusion to facilitate the coder's observation of facial muscle movements during the ME coding.


\subsubsection{EMG Signal Pre-Processing}
\begin{figure*}
  \centering
\subfloat[]{\includegraphics[width=0.19\textwidth]{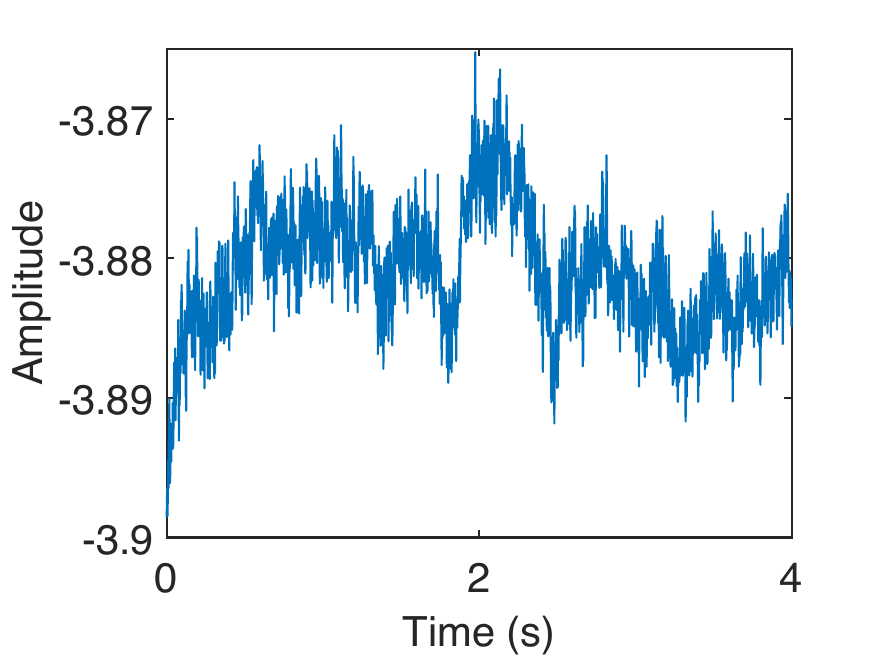}\label{fig:emgPre1}}\hfil
\subfloat[]{\label{fig:emgPre2}\includegraphics[width=0.19\textwidth]{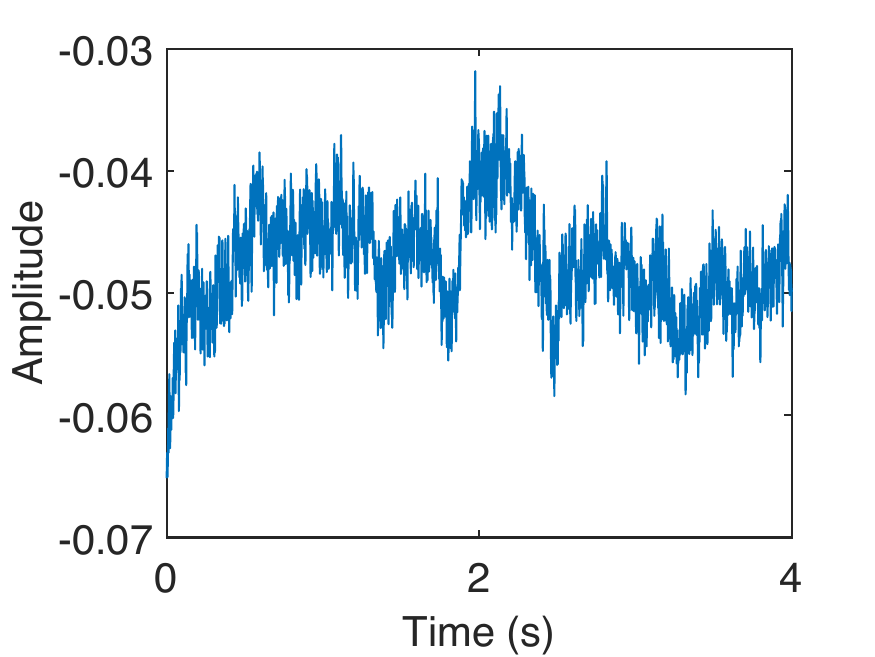}}\hfil
\subfloat[]{\label{fig:emgPre3}\includegraphics[width=0.19\textwidth]{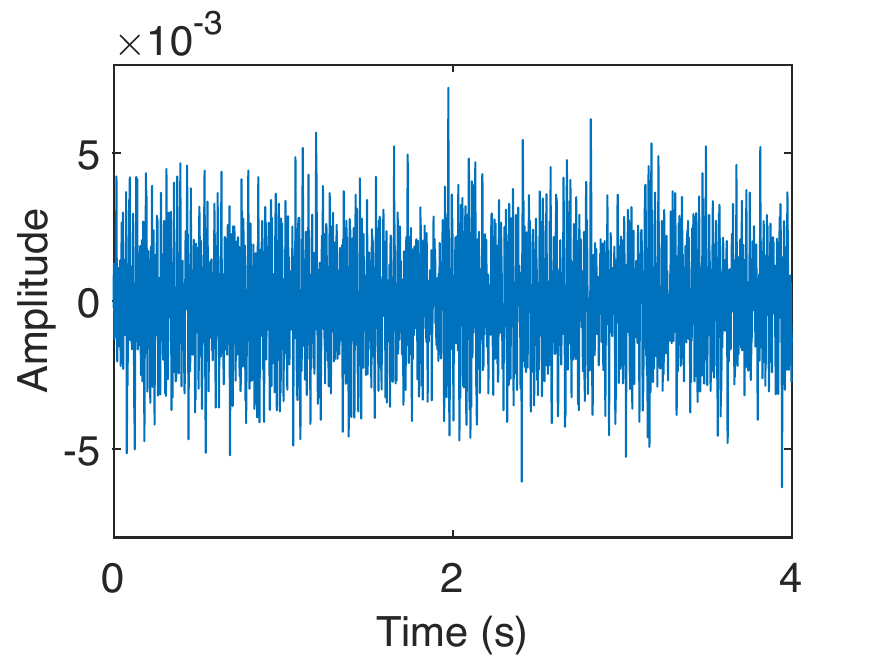}}\hfil
\subfloat[]{\label{fig:emgPre4}\includegraphics[width=0.19\textwidth]{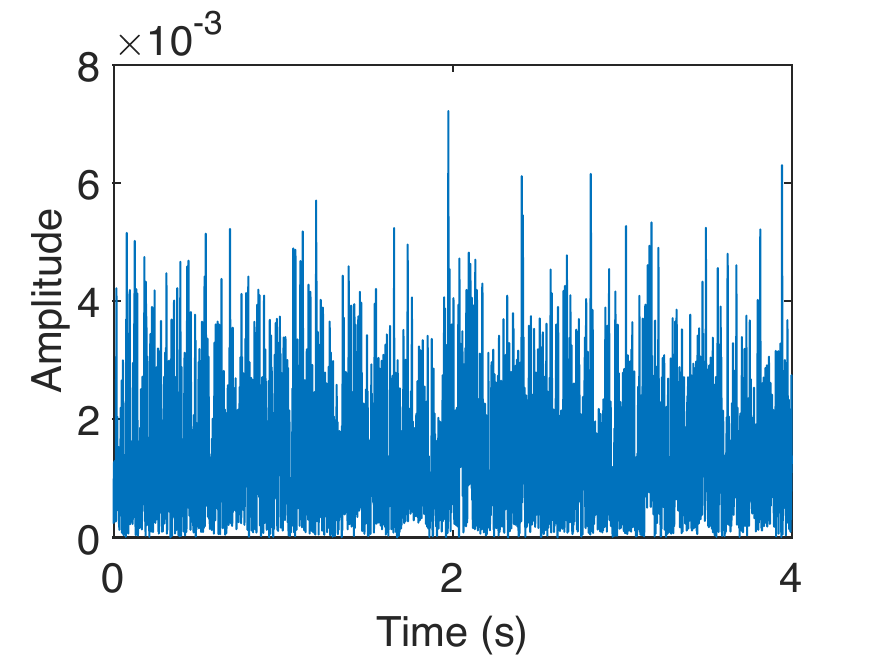}}\hfil
\subfloat[]{\label{fig:emgPre5}\includegraphics[width=0.19\textwidth]{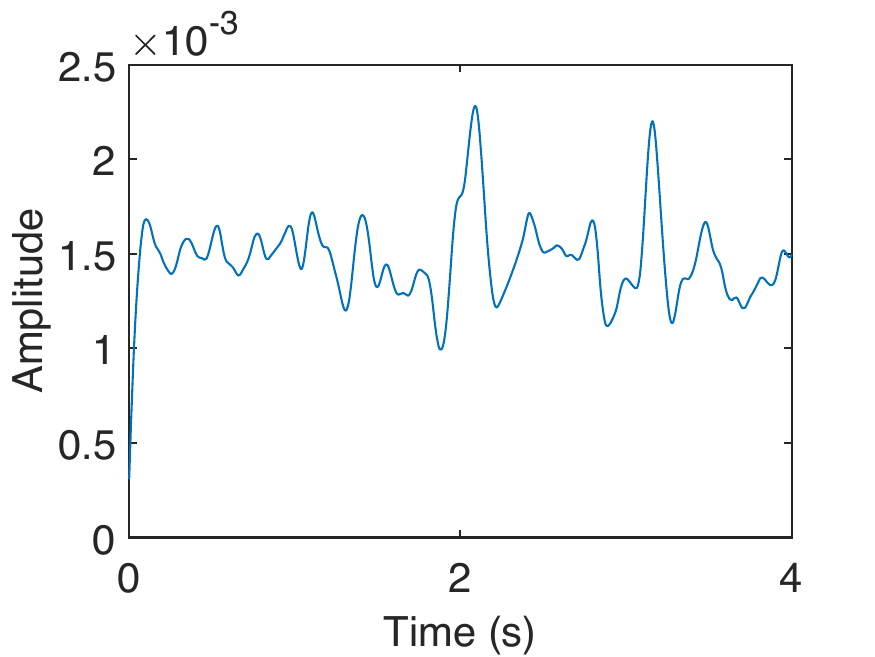}}\hfil
  \caption{Four steps of EMG signal pre-processing: removing DC offset, denoise, full-wave rectification, and linear envelope. (a) Raw EMG data; (b) EMG with DC Offset Removed; (c) EMG with Noise Removed; (d) EMG with Full-Wave Rectification; (e) Linear Envelope of EMG.}
  \label{fig:pre-processing}
\end{figure*}
The raw EMG signal was pre-processed before the meaningful interpretation of the EMG data because the raw EMG signals usually carry a lot of noise. In our study, the EMG signal pre-processing involved removing Direct Current (DC) offset, denoise, full-wave rectification, and linear envelope, as shown in Fig.~\ref{fig:pre-processing}. All of the processes were implemented on the MATLAB platform. 
\par
\textbf{Removing DC offset}: The DC component represents the mean value of the signals. By subtracting the mean value from the raw EMG signal $E_r$, the EMG signal with DC offset removed $E_d$ was obtained (Fig.~\ref{fig:emgPre2}).
\begin{equation}
 E_d = E_r - \text{mean}(E_r)
\end{equation}
where $\text{mean}(\cdot)$ denotes calculating the average value of the variable.

\par
\textbf{Denoise}: The predominant frequency range of facial EMG signals is 20-500 Hz~\cite{van2010facial}. Thereby, to minimize noise within the EMG signals, we applied a 2nd-order band-pass Butterworth filter (BF$_{\text{bp}}$) from roughly 20 Hz to 450 Hz on $E_d$ and obtained $E_b$ (Fig.~\ref{fig:emgPre3}). 
\begin{equation}
 E_b=  \text{BF}_{\text{bp}}(E_d, 2, [Wn_{1},Wn_{2}])
\end{equation}
where $Wn_{1}$ and $Wn_{2}$ represent the lower and the higher cutoff frequencies, respectively, $Wn_1 =  20\times 2/1000$, $Wn_2 =  450\times 2/1000$.

\par
\textbf{Full-wave rectification}: The EMG signal fluctuates around the 0-value baseline. Thus, the EMG signal $E_b$ was rectified by absolute value to achieve full-wave rectification of the EMG signals~\cite{hess2009facial} (Fig.~\ref{fig:emgPre4}). 
\begin{equation}
 E_f=  \text{abs}(E_b)
\end{equation}
where $\text{abs}(\cdot)$ denotes calculating the absolute value of the variable.

\par
\textbf{Linear envelope}: A more intuitive data representation could facilitate researchers to compare and analyze the amplitude and wavelength of the EMG signals. The linear envelope of the EMG signals $E_l$ is implemented through a low-pass filter which is a 2nd-order Butterworth filter (BF$_{\text{lp}}$) with a low-pass cutoff frequency of 6 Hz (Fig.~\ref{fig:emgPre5}).
\begin{equation}
 E_l=   \text{BF}_{\text{lp}}(E_f, 2, Wn)
\end{equation}
where $Wn=  6\times 2/1000$ denotes the normalized cutoff frequency.
All the following data analysis is based on the envelope EMG signal $E_l$.

\subsubsection{EMG Indicators}
Two indicators of EMG signals, MVC$\%$ and iEMG, are chosen for the quantification of ME external characteristics.
\par
\textbf{MVC$\%$}: For each facial expression, the EMG signal of the channel with the most distinct signal variation at the appearance of the expression is conserved as the specific EMG signal for this facial expression. 
     $$
     A_{i,j}^{k} = \max(\text{Amp}(E_{t,i,j}^{k})
     $$
     \begin{equation}
  E_{i,j} =E_{t,i,j}^{k} \quad \text{where} \quad k = \argmax_{k \in \{1, \dots, 7\}} \left( A_{i,j}^{k} \right)
  \end{equation}
 where $i$, $j$, and $k$ denote the index of participants, the expression, and the EMG channels, respectively. 
    Then, as shown in Eq.~\ref{eq:mvc}, the amplitude peak value $A_{i,j}^{k}$ of the EMG signal $E_{i,j}$ of each facial expression was divided by the MVC of the corresponding channel to obtain the amplitude ratio. This ratio, denoted as MVC$\%$, represents the degree of muscle activation. This relative comparison could minimize between-participant variability~\cite{lehman1999importance}.
    \begin{equation}
    \label{eq:mvc}
    \text{MVC\%}_{i,j}^{k} = \frac{A_{i,j}^{k}}{ \text{MVC}_{i}^{k}}
    \end{equation}
    
   
    \par
    \textbf{iEMG}: The integrated ENG (iEMG) represents the mathematical integration of the absolute value of the EMG signal $E_{i,j}$. In other words, this indicator measures the sum of muscle fibers' temporal and spatial electrical activity in the recorded electrode ares~\cite{moritani1982electromyographic}. Furthermore, as shown in Eq.~\ref{eq:iemg}, the EMG signals were normalized, enabling the iEMG result comparison across participants. 
  \begin{equation}\label{eq:iemg}
       \text{iEMG}_{i,j} = \int_{t_1}^{t_2} |\text{Norm}(E_{i,j}(t))| dt
   \end{equation}
   where $t_1$ and $t_2$ represent the begin and ending time of facial expression, respectively. Norm denotes the normalization calculation.

\subsection{Self-report-based Internal Characteristic Measurement}


In our study, we employed the second-generation ME elicitation paradigm~\cite{li2022casme3}, using intense emotional stimuli to evoke participants’ emotions, as described in Section~\ref{sec:elicitation}. Participants were instructed to maintain a neutral facial expression while viewing these stimuli. Video stimuli often trigger sensations, perceptions, and emotions, leading to facial expressions. We assessed participants’ awareness and control over these expressions using two self-report methods.
First, while viewing the stimuli, the participants were required to press the `L' on the keyboard when they were aware of an expression leakage. The keystroke response responds to the rate of self-awareness. Subsequently, \highlight{after viewing each video clip, participants were asked to assess their ability to control their facial expressions using a 9-point Likert scale. The scale ranges from 0 (no control over facial expressions) to 8 (complete control), allowing participants to select the number that best represents their level of self-control during the video~\cite{tangney2018high}.} Using the two self-report methods, we could analyze whether there is a significant difference between the occurrence of MEs and MaEs.

\section{Experiment}
\subsection{Participants}
Thirty-five participants (17 females; Mean age  = 24.7 years, Standard Deviation = 1.6 years) participated in the experiment. They all had normal or corrected-to-normal vision. In addition, none of them had a history of neurological, psychiatric, or other serious illnesses that could affect the experiment. Each participant was given a detailed description, signed an informed consent form before the experiment, and paid after finishing the experiment. Our study followed the Declaration of Helsinki and was approved by Institute of Psychology, Chinese Academy of Sciences.

\subsection{Experiment Scene }
All equipment was arranged in a quiet, soundproof lab with stable, soft lighting provided by two LED lights with reflective umbrellas. A monitor with 2564 $\times$  1440 pixels resolution and speakers was placed in front of the participants to play the stimuli. A Logitech C1000e camera positioned on the monitor recorded the participants’ facial responses at a resolution of 3840 $\times$  2160 pixels and 30 FPS. The captured facial areas had a resolution of approximately 700 $\times$ 900 pixels.
As shown in Table~\ref{tab:ddbpixel}, the SAMM database~\cite{davison2018samm} and MMEW~\cite{ben2021video} have the highest facial region resolution, around 400 x 400 pixels, among published ME databases. Our facial region resolution is four times higher, enabling more accurate coding of facial expressions. \highlight{The camera’s 30 FPS frame rate effectively captures facial movements at intervals above 33 milliseconds, which suffices for our purposes. In our study, video recordings of MEs corroborate EMG data, helping us identify and match the timing of expressions with corresponding EMG signals. EMG signals sampled at 1000 Hz are then used to precisely quantify the temporal characteristics of MEs.}
\begin{table}[b]
\begin{center}

    \caption{Facial region resolution summary of public spontaneous ME databases. }\label{tab:ddbpixel}
     \centering
   
    \begin{tabular}{|c|c || c | c|}
  \hline
     Databases&  Resolution & Databases&  Resolution\\ \hline
         CASME~\cite{yan2013casme}& 150 $\times$ 190&     SMIC~\cite{li2013spontaneous}& 190 $\times$ 230 \\
         CASME~\uppercase\expandafter{\romannumeral2}~\cite{yan2014casme} &  280 $\times$ 340& CAS(ME)$^2$~\cite{qu2017cas}&  200 $\times$ 340\\
         SAMM~\cite{davison2018samm}&   400 $\times$ 400&          MMEW~\cite{ben2021video}&  400 $\times$ 400\\
         CAS(ME)$^3$~\cite{li2022casme3} &  250 $\times$ 300 & 4DME~\cite{li20224DME} & 160 $\times$ 200\\ \hline
         
    \end{tabular}

\end{center}
\end{table}
\par

\subsection{Elicitation Materials}\label{sec:elicitation}

Numerous methods are commonly used to elicit participants’ emotions, with video elicitation offering advantages such as dynamism and non-deceptiveness~\cite{gross1995emotion}. We selected 18 storytelling video clips from two databases that are frequently used to elicit expressions and micro-expressions (MEs). Specifically, 12 clips were ME elicitation stimuli from the CASME database~\cite{yan2013casme,yan2014casme}, and six were chosen from the Chinese Positive Emotion Database (CPED)\cite{zhang2021cped}. To enhance the emotional variety, six additional open-source clips from films were added. In total, as listed in Table\ref{tab:elicitation}, there are 24 video clips, with four clips representing each of the six emotion categories: happiness, sadness, disgust, anger, fear, and surprise~\cite{donia2014spontaneous}.
\begin{table*}[h]
\begin{center}
 \caption{Video clips selected as elicitation stimuli for this experiment.}\label{tab:elicitation}
     \centering
 \begin{tabular}{|c|c|c|c || c|c|c|c|}
  \hline
     Clip Name &  Source & Emotion & Time & Clip Name& Source & Emotion& Time\\ \hline
        Massacre& CASME  & Anger& 144s & School bullying & CPED  & Anger& 220s  \\
        Dog abuse & CASME  & Anger& 96s & Attack & Film & Anger& 144s\\
        Accident & CASME & Sadness& 117s & CJ7 &  CASME & Sadness& 140s \\
        Farewell & CPED  & Sadness& 159s & Olympic Games & CASME  & Sadness& 91s \\
        Eat Cockroach & CASME  & Disgust& 42s & Young emperor & CPED  & Disgust& 145s \\
        Eat Worm & CASME & Disgust& 34s & Myopia operation & CASME  & Disgust & 116s \\
        Spider & Film & Fear& 25s & Toilet psychic incident &  CPED & Fear& 68s \\
        Elevator thriller & CASME & Fear& 77s &  Teiko & CASME  & Fear & 161s\\
        Climb the window  & Film & Surprise &67s  & War &  Film & Surprise& 51s \\
        Magic & Film & Surprise& 53s &  Car chase &  Film & Surprise& 35s\\
        Honey-trap  & CASME  & Happiness& 67s &  Cartoon & CASME  & Happiness & 92s\\
        Vocal concert   & CPED  & Happiness& 172s & Won the game & CPED  & Happiness& 90s\\\hline         
    \end{tabular}
\end{center}
\end{table*}

\subsection{Procedure}

\begin{figure}[b]
  \centering
  \includegraphics[width=\linewidth]{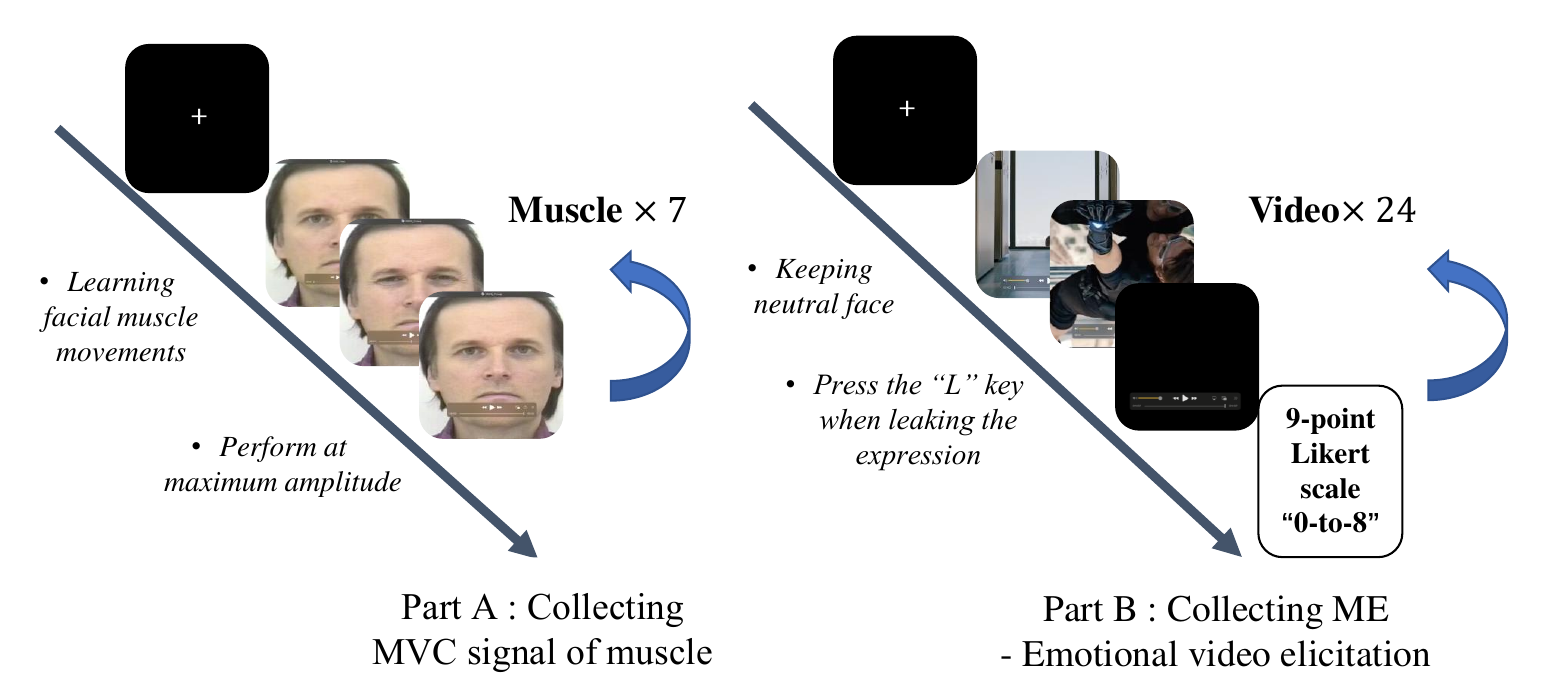}
  \caption{The experimental procedure consists of two parts, A and B. Part A collected the MVC of each muscle, and part B used video stimuli to evoke expressions.}
  \label{procedure}
\end{figure}

The experimental procedure, illustrated in Fig.~\ref{procedure}, began with a detailed explanation of the study’s requirements to the participants, followed by the signing of an informed consent form. During the preparation phase, electrodes were attached to the participants’ faces for facial EMG acquisition in parts A and B.

\par
In part A, participants were guided to perform seven different facial muscle movements corresponding to the seven EMG channels. Each movement was repeated three times at maximum intensity to capture the MVC on the seven facial muscles, which served as a baseline for further analysis.
\par
In part B, we elicited MEs using 24 emotional video stimuli, presented in a randomized, balanced order to minimize stimulus interaction. These clips, lasting 30 to 120 seconds each, were evenly distributed across six emotion categories. Participants were instructed to maintain a neutral expression while watching the stimuli to encourage the emergence of MEs when trying to hide genuine expressions. Additionally, participants were asked to press the ‘L’ key if they perceived an expression leak. After viewing each video, they completed a 9-point Likert scale to rate their self-control over their emotions.


\subsection{Facial Expression Coding}
\begin{figure}[b]
    \centering
    \includegraphics[width = \linewidth]{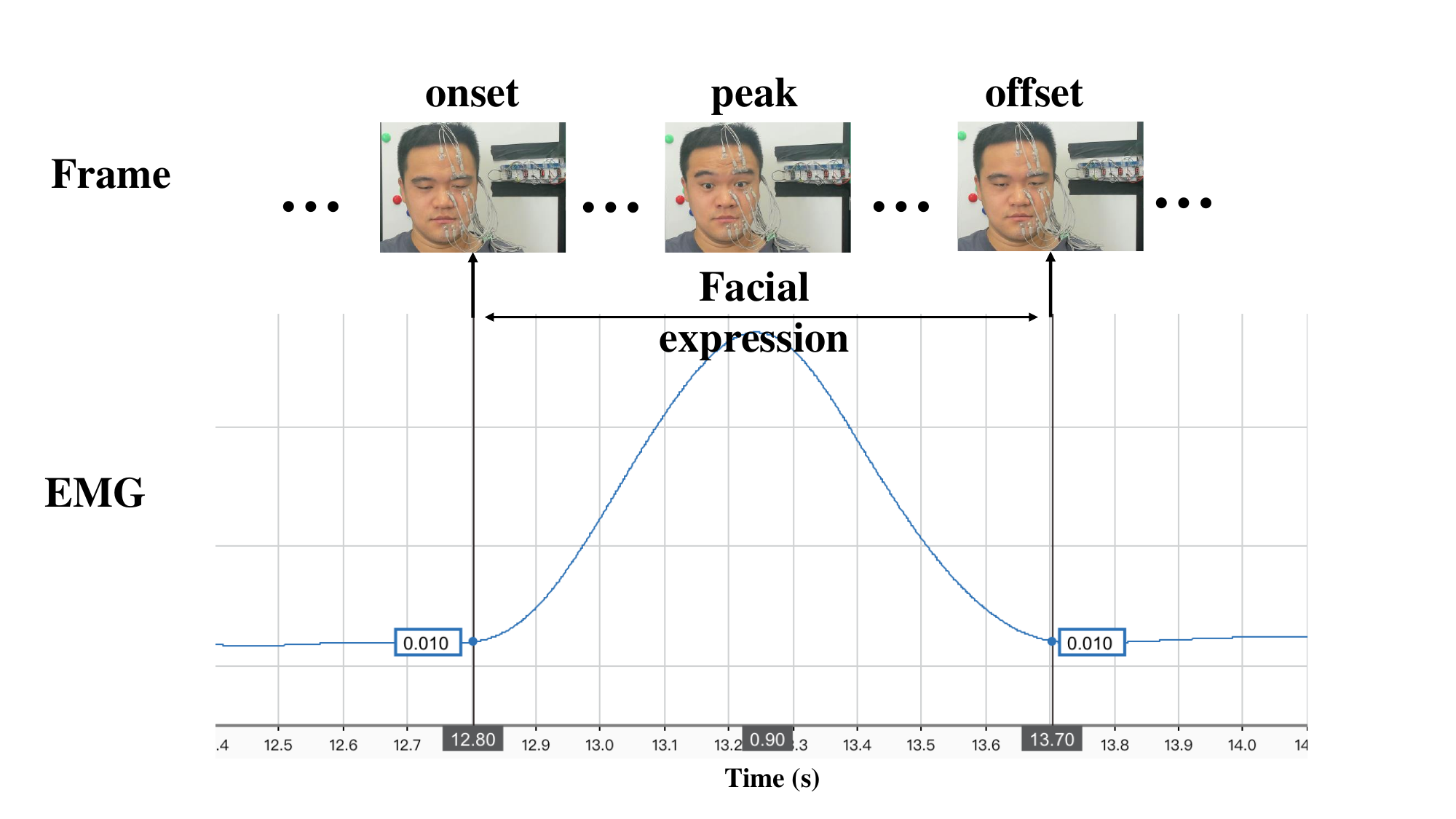}
    \caption{EMG signal extraction based on the frame-based annotation of facial expression}
    \label{fig:label}
\end{figure}

To identify the EMG signals corresponding to facial expressions, we first coded the expressions in the recorded videos frame by frame using the Facial Action Coding System (FACS). We then searched for distinct EMG signal amplitudes within the time intervals corresponding to the onset and offset frames of the coded expressions.
Specifically, one trained coder manually annotated the videos, marking the onset, apex, and offset of all emotional expressions while excluding facial movements caused by mechanical actions like swallowing and blinking. After coding, we synchronized the timing of these facial expressions with the EMG signals using the digits displayed on the digital tube.
As illustrated in Fig.~\ref{fig:label}, if the expression’s occurrence coincided with a variation in the EMG signal, the channel exhibiting the most significant change was selected for further analysis as the EMG signal representing that expression.

\section{Result and Discussion} \label{sec:result}
n the experiment, 406 facial expressions were initially coded, but 26 were excluded during data processing. Specifically, 21 expressions were removed due to the absence of corresponding EMG signals, likely because only seven facial muscles on the left side were monitored, leaving other facial regions uncaptured. Additionally, 5 expressions were excluded due to the failure in capturing the maximal voluntary contraction in some participants, preventing cross-participant and cross-channel comparisons. Consequently, 380 expressions with corresponding EMG signals were included in the analysis, comprising 233 MEs and 147 MaEs.

\subsection{External Characteristics} \label{subsec:rstEC}
We analyzed the data of EMG signals to explore the two external characteristics of MEs: low intensity and short duration. 

\subsubsection{Low Intensity}

\begin{figure}[b]
  \centering
  \includegraphics[width=0.8\linewidth]{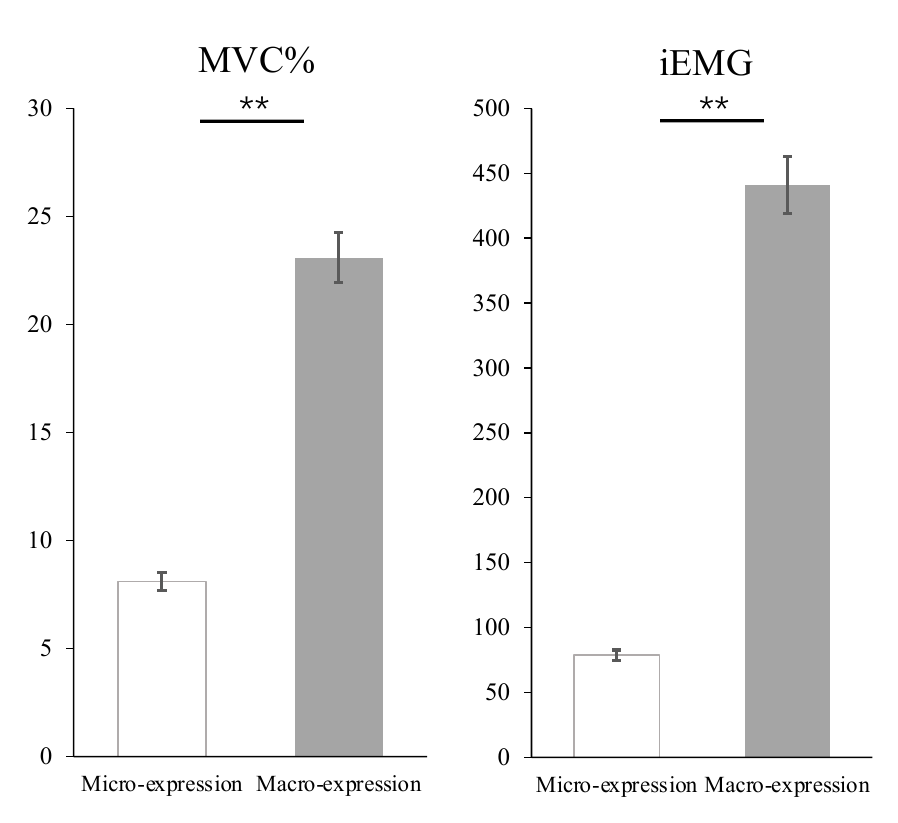}
  \caption{Statistical analysis on MVC$\%$ and iEMG between MEs and MaEs. The EMG data of MEs and MaEs were significantly different by independent sample T-test with a 95$\%$ confidence interval. The notation ** represents \textit{p} $< 0.01$.}
  \label{fig:result}
\end{figure}

MVC$\%$ was used as an EMG indicator to measure the intensity characteristics of expressions. The basic statistics of the MVC$\%$ are listed in Table~\ref{tab:number1}. We analyzed each facial expression using Independent Samples T-test analysis to explore whether there were differences in intensity under the ME and MaE. The results showed that the intensity differed significantly between the two groups, reflected in the left side of Fig.~\ref{fig:result}. MVC$\%$ was significantly lower for MEs ($M=8.11$, $SD=8.54$) than for MaEs ($M=23.09$, $SD=21.27$), $t(378)=9.6$, $p<0.05$, $d=1.01$. MVC$\%$ of MEs was lower than MaEs by 1.01 times the standard deviation. According to Cohen's criterion, the effect size could be considered to be at a high level.

\begin{table}[b]
\begin{center}
\caption{The numerical distribution of the two indicators of ME and MaE. \textit{N}, \textit{Mean} and \textit{SD} represent the number of facial expressions,  the average value, and the standard deviation, respectively.
In addition, in the results of the T-test, \textit{df}, \textit{t}, \textit{p} and \textit{d} represents the degree of freedom, the test quantity, the error probability that the results are representative of the population, and the standard difference between the mean values, respectively.}
  \label{tab:number1}
  \centering
\resizebox{\linewidth}{!}{
\begin{tabular}{|c|c|c|c|c|c|c|c|c|}
\hline
   Indicators & Type &\textit{N}  & \textit{Mean} &\textit{SD}  &\textit{df}  &\textit{t}  &\textit{p}  &\textit{d} \\\hline
\multirow{2}{*}{MVC\% }& MaE      & 147 & 23.09 & 21.27& \multirow{2}{*}{378} & \multirow{2}{*}{9.6}& \multirow{2}{*}{0.00}& \multirow{2}{*}{1.01}\\
           & ME     & 233 & 8.11 & 8.54  &&&& \\
\multirow{2}{*}{iEMG }      & MaE      & 147 & 440.89 & 459.97 & \multirow{2}{*}{378} & \multirow{2}{*}{11.89}& \multirow{2}{*}{0.00}& \multirow{2}{*}{1.25}   \\
           & ME      & 233 & 78.8  & 55.6 &&&&    \\ \hline
\end{tabular}}

\end{center}
\end{table}

\par

MVC$\%$ provides feasibility for directly comparing muscle activities across channels and participants. The average MVC$\%$ apparently demonstrates the characteristics of low intensity. Furthermore, the intensity significantly differs between ME and MaEs. 
Hence, the minor MVC\% value of ME could be considered as a preliminary empirical basis for ME characteristics of low intensity.

\par

\begin{figure}[t]
  \centering  \includegraphics[width=\linewidth]{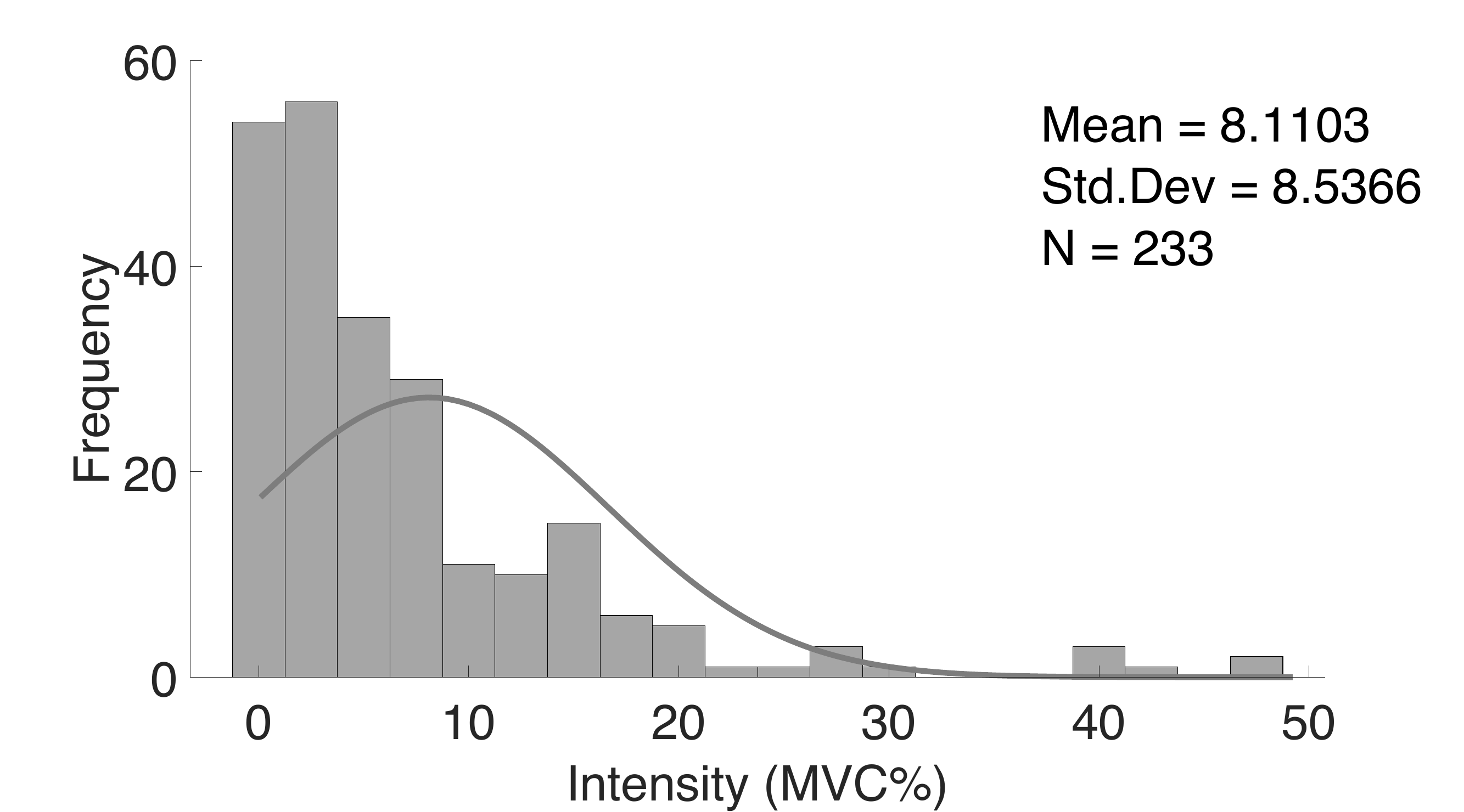}
  \caption{Distribution of ME intensity}
  \label{fig:intensity}
\end{figure}

In addition to the above qualitative observations, we also inferred EMG-based interval estimates. First, the histogram visualizes the intensity distribution (Fig.~\ref{fig:intensity}). Then, we used the interval estimation method in statistics to infer the intensity intervals of the population parameters of MEs. As listed in Table~\ref{tab:intervalEstimation}, the results of the interval estimation at the 95$\%$ confidence interval showed that the lower confidence limit of the intensity indicator MVC$\%$ of MEs was 7.01$\%$, and the upper confidence limit was 9.21$\%$, which means that the intensity of MEs was roughly between 7$\%$ and 9.2$\%$ of the MVC.

\begin{table}[b]
\begin{center}

\caption{Interval range of duration and intensity of ME and MaE obtained by Interval Estimation}
\label{tab:intervalEstimation}
  \centering
\begin{tabular}{|c|c|c|c|}
\hline
 Characteristics &  Type &  Lower limit  & Upper limit \\\hline
 \multirow{2}{*}{Intensity (MVC\%) }& MaE     & 19.63 & 26.56\\
           & ME     & 7.01 & 9.21    \\
\multirow{2}{*}{Duration (Second)}  & MaE & 1.027 & 1.295\\
           & ME      & 0.307 & 0.327  \\ \hline
\end{tabular}
\end{center}
\end{table}

\subsubsection{Correlation between Duration and Intensity}

iEMG, as an EMG indicator with two dimensions of time and amplitude, represents the normalized integrated EMG value. Independent Sample T-test results showed that iEMG of ME ($M=78.8$, $SD=55.6$) was significantly lower than that of MaE ($M=440.89$, $SD=459.97$), $t(378)=11.89$, $p<0.05$, $d=1.25$. 
iEMG for MEs was 1.25 times lower than that for MaEs in standard deviation. 
According to the Cohen criterion, the effect size could be considered to be at a high level. As shown on the right side of Fig.~\ref{fig:result}, the intensity and duration in the EMG signal of MEs were significantly lower than that of MaEs. Since the iEMG reflects a composite character of the amplitude and time of EMG signal, the low intensity and short duration of MEs can be tentatively verified. 
\par
\begin{figure}[t]
  \centering
  \includegraphics[width=0.8\linewidth]{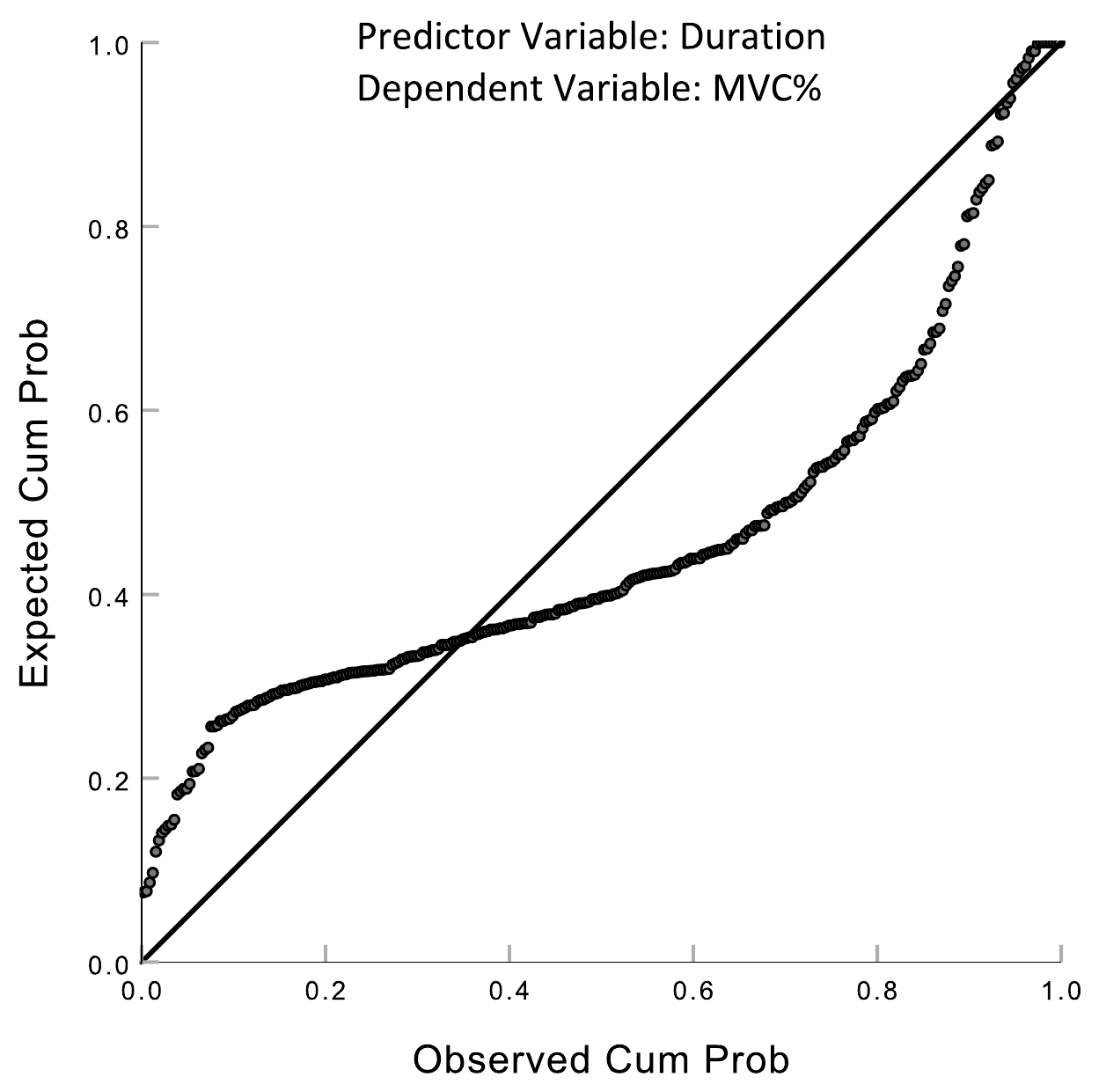}
  \caption{Normal P-P Plot of Regression Standardized Residual. 
  }
  \label{fig:related_result}
\end{figure}

Moreover, we explored the relationship between two characteristics of expressions: intensity and duration. Statistical analysis were conducted with intensity as the dependent variable and duration as the predictor variable. Linear regression analysis shown in Fig.~\ref{fig:related_result} revealed that duration was a significant predictor variable of intensity, $\beta = 0.492$, $t(378) = 10.98$, $p < 0.05$. The experimental results indicate a significant linear correlation between the duration and intensity of MEs; specifically, a shorter duration corresponds to a lower intensity. This result highly coincides with the results of the iEMG indicator.

\subsubsection{Short Duration}
As previously described, our analysis of the correlation between intensity and duration, along with the iEMG indicator that reflects both temporal and intensity characteristics, leads us to conclude that the ME duration is brief, significantly shorter than that of MaE. This forms the basis for our further investigation into the ME duration.
\begin{figure}[b]
  \centering  \includegraphics[width=0.9\linewidth]{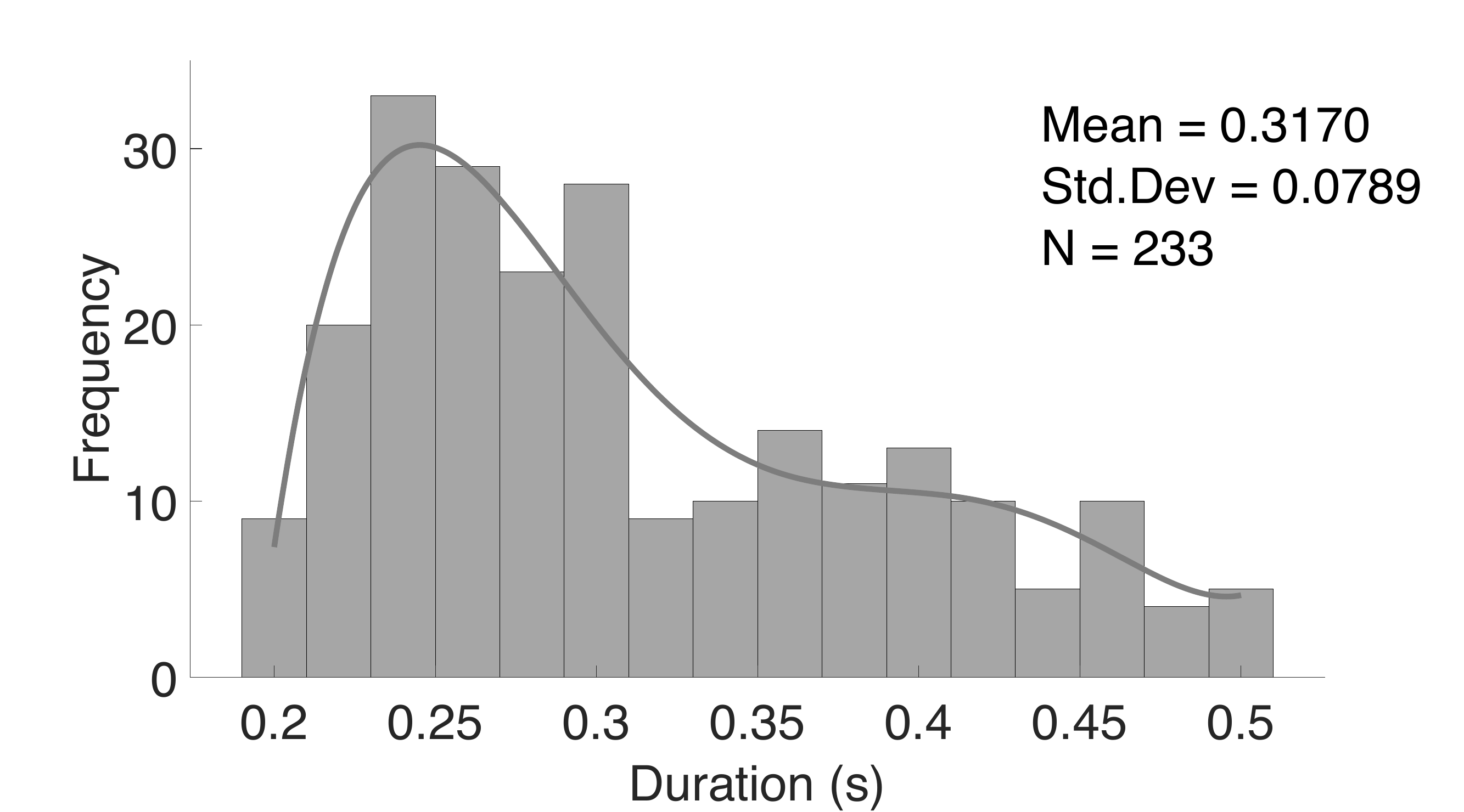}
  \caption{Distribution of ME duration 
  }
  \label{fig:duration}
\end{figure}

The duration distribution histogram is illustrated Fig.~\ref{fig:duration}. Similar to the intensity estimation, we also performed an interval estimation of the duration of MEs. The results showed that the lower confidence limit for the ME 
 duration was 0.307 s, and the upper confidence limit was 0.327 s, with a confidence probability of 95$\%$. This interval estimation could be representative of the population parameters behind the ME samples used for duration analysis. Our results for ME duration interval estimation are consistent with the most commonly used definition of 1/2 s by current researchers ~\cite{ekman2005face,polikovsky2009facial}. As listed in Table~\ref{tab:intervalEstimation}, we also obtain interval estimates of the duration and intensity of MaEs.
 Overall, by taking advantage of EMG, we inferred the main distribution intervals of two external characteristics for MEs, which lays the foundation for future research on the physiological mechanisms.

\par
\begin{figure}[t]
    \centering
    \includegraphics[width= \linewidth]{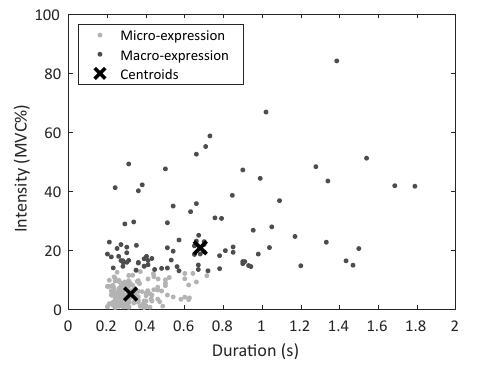}
    \caption{\textit{k}-means clustering for intensity (MVC\%) and duration. The 380 facial expressions were divided into two clusters. The crossed symbols in the figure represent the centroids of these clusters, aligning with the distribution and interval estimates of duration and MVC\%. For better visualization of sample distribution, facial expressions lasting longer than the mean plus variance in each category were excluded from the figure.}
    \label{fig:cluster}
\end{figure}

In order to further verify the reliability of the duration and intensity characteristic intervals, we used \textit{k}-means cluster for an unsupervised analysis on 380 expression samples, i.e., the expression samples were not classified as ME and MaE based on the 500 ms threshold in clustering analysis. 
Specifically, \textit{k}-means~\cite{lloyd1982least} algorithm divided 380 indicator combinations of duration and intensity into two clusters. The positions of the centroids in two clusters were optimized by the iterations calculating the indicator-point-to-cluster-centroid distance. 
As shown in Fig.~\ref{fig:cluster}, the clustering result showed that the intensity and duration of one of the categories were within the above range of the psychological-based interval estimation of ME shown in Fig.~\ref{fig:intensity} and Fig.~\ref{fig:duration}. This consistency indicates that MEs and MaEs can be distinguished just as well when departing from the 1/2~s division criterion. Thus, through statistical analysis, psychological analysis, and clustering calculations, our study shows that MEs can be effectively distinguished from MaEs by intensity and duration. This conclusion enriches the previous view that duration is the only difference between MEs and MaEs~\cite{shen2012effects}.
\par

Apart from analyzing ME duration in our experiment, we also explored the definition of ME duration in previous studies. While many definitions exist, our interval estimation aligns closely with Ekman and Polikovsky~\cite{ekman2005face,polikovsky2009facial}. We addressed the inconsistencies in other results, noting that the process of ME production differs fundamentally from ME recognition. For instance, definitions like 1/25~s and 1/5~s~\cite{ekman1975unmasking}, derived from recognition methods, are not suitable for defining the actual duration of ME production. MEs, as theoretically defined, are brief, involuntary leaks of genuine emotion, reflecting the cognitive state at the moment of occurrence. Meanwhile, the extremely small value of 1/25~s, likely reflects the minimum threshold for visual recognition by the human eye~\cite{becattini2022understanding}, rather than a meaningful duration of ME occurrence.
The significance of our precise interval is to define ME duration with a 95\% confidence level of high occurrence probability, unlike the vague temporal definitions used before. 

\par
Furthermore, we discuss the commonly used ``less than 1/2 s'' definition in computer science, highlighting the lack of empirical evidence from cognitive neurology to support it. Without concrete neural or physiological markers, the boundary for ME duration is behaviorally derived: the shorter the duration, the more likely an expression is to be a ME. However, the broad 1/2~s boundary is more practical for computer-based ME analysis, which requires large sample sizes.

\par
In summary, previous studies relied solely on duration to define MEs, without considering intensity~\cite{ekman1969nonverbal,ekman1975unmasking,ekman1985telling,ekman2005face}. We demonstrated significant differences between MEs and MaEs in both duration and intensity. Our study proposes intensity as an additional criterion, offering a more precise definition of MEs. This two-dimensional approach of duration and intensity serves as a foundation for intelligent ME analysis, enabling more targeted machine learning methods and improving the distinction between MEs and other facial movements in practical applications.

\subsection{Internal Characteristic}

The external appearance of MEs occurs only after the individual experiences emotional reactions at the cognitive level. Therefore, after analyzing the external characteristics of MEs, we attempted to explain the internal characteristics through the rate of awareness and control, via keystroke response and self-reports. 
In particular, Chi-square tests were conducted separately for the frequencies of keystroke response and control rate reported in the self-reports. 
\par
Besides, we would like to make a distinction between the two terms in order to avoid confusion. In the context of this paper, unawareness represents the ability of the individual's senses to be unable to perceive, while unconsciousness is the state of the individual's inability to mental percipience. 
\begin{figure}[b]
  \centering
  \includegraphics[width=0.8\linewidth]{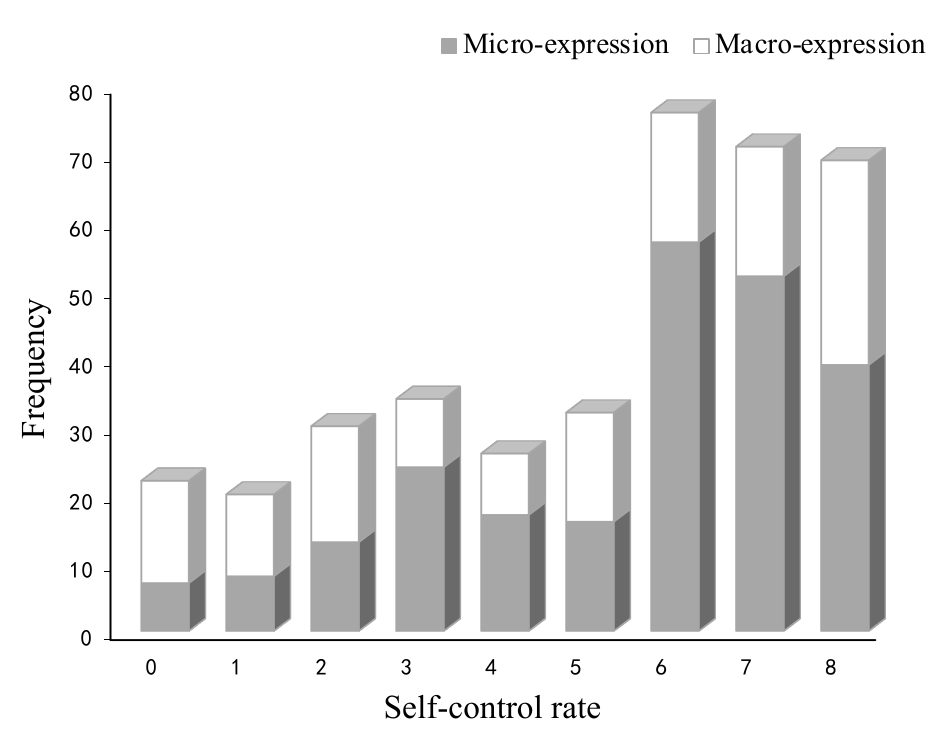}
  \caption{The facial expression frequencies under different self-control rates}
  \label{fig:control}
\end{figure}

\subsubsection{Self-Awareness}
The types of expression (ME or MaE) and the keystroke response (yes or no) were analyzed for correlation. The basic statistics of the keystroke frequency are shown in Table~\ref{tab:key}. There was a significant correlation between the type of expression and the ability to be aware of the expression, $\chi^{2}(1, N=380)=26.54$, $p<0.05$, \textit{Cramer's V} = 0.26. 
More expressions reported by individuals were MaEs (70\%). On the contrary, most of the MEs were not reported (93\%).
This result indicated that individuals were more likely to be aware and report MaEs by keystrokes and less likely to be aware when MEs appear. 
\par
This finding suggests that ME leakage is neither consciously aware nor consciously expressed~\cite{warren2009detecting}. Furthermore, based on our experimental results and previous research, it can be assumed that duration can serve as a predictor of intensity, while intensity can predict self-awareness~\cite{qu2017you}. Thereby, as a facial movement with short-duration and low-intensity characteristics, MEs occur below the threshold of self-awareness.

\begin{table}[t]
\caption{The numerical distribution of self-awareness and facial expression frequency. The response with a keystroke is represented as aware, and the response without a keystroke is represented as unaware. ME and MaE represent micro- and MaEs, respectively.}
  \label{tab:key}
   \centering
   
        \begin{tabular}{|c|c|c|c|}
        \hline
         Keystroke response  & ME  & MaE  \\ \hline
        Aware(Yes)     & 17 (7$\%$) & 39 (27$\%$)  \\
        Unaware(No)    & 216 (93$\%$) & 108 (73$\%$) \\
        Total (Frequency)      & 233 (100$\%$) & 147 (100$\%$) \\
        \hline
        \end{tabular}
    
\end{table}

\subsubsection{Self-Control}
The correlation of expression type (ME or MaE) and the emotional self-control rate (0-8) were also explored. In the chi-square test, $\chi^{2}(8, N=380)=30.06$, $p<0.05$, \textit{Cramer's V} = 0.28. The results showed a significant correlation. Precisely, the higher the self-control rate, the more MEs appear, as shown in Fig.~\ref{fig:control}. Individuals leaked plenty of MEs when they perceived they had higher self-control over their emotions. Therefore, self-control over emotions could only be reflected in the restriction of MaEs. MEs were challenging to control and were spontaneous.
\par
This finding is corroborated by our statistical analysis, which showed that MaEs are less likely to leak and are more easily suppressed compared to MEs. This result aligns with earlier hypotheses suggesting that individuals’ genuine emotions manifest uncontrollably as MEs, appearing briefly and unnoticed on their faces~\cite{haggard1966micromomentary,ekman1985telling,ekman2006flawed}. In other words, the leakage of MEs is unconscious and beyond the individual’s control. Moreover, the results support the idea that MEs can reveal an individual’s true emotions. This conclusion justifies the use of MEs as a valid indicator in practical applications and enhances the reliability of ME analysis in real-world settings.

\section{\highlight{CASMEMG dataset}}
\highlight{We organized the data from our previous study to develop a ME database, CASMEMG. This database contains 147 MaE and 233 ME samples, each accompanied by corresponding EMG signals and videos. The original raw data from which these samples were segmented is also provided. In this section, we introudce the annotation of emotion labels and provide a baseline method and results for EMG-based facial expression interval detection. The CASMEMG is open for application, with the aim of promoting the study of micro-expression mechanisms and the dynamics of movement changes grounded in EMG signals.}

\subsection{\highlight{Emotion Annotation}}

\highlight{We have coded the samples in the CASMEMG dataset for Action Units (AUs) and emotions. In particular, the database includes 380 facial expression samples with both EMG and video data. All samples have been coded for AUs and emotions by two trained coders, exhibiting high coding consistency. Initially, the AU coding followed the requirements and guidelines of the FACS system. AUs that have emotional significance in the facial expressions were coded, covering 27 AUs related to facial expressions. Based on the FACS coding method, the onset frame, apex frame, and offset frame of the 380 expressions, along with their corresponding AU combinations, were determined.}
\par
\highlight{After the two coders independently completed the coding, the consistency $R_C$ of the AU coding was calculated. The method referred to the reliability calculation formula used in CAS(ME)$^2$~\cite{qu2017cas}: }
\begin{equation}
    \highlight{R_C=\frac{N_{C_1 \cap C_2}\times 2}{N_{C_1 \cup C_2}}}
\end{equation}
\highlight{where $N_{C_1 \cap C_2}$ and $N_{C_1 \cup C_2}$ represent the number of samples in the intersection and union of the coding results from two coders, respectively.}
\highlight{In the database created for this study, the consistency of AU coding by the two coders was 83\%, indicating high reliability. Additionally, each AU was coded for whether it was unilateral/bilateral, and it was found that only one expression had inconsistent coding, with a consistency as high as 99.7\%. }
\par
\begin{figure}
    \centering
    \includegraphics[width=\linewidth]{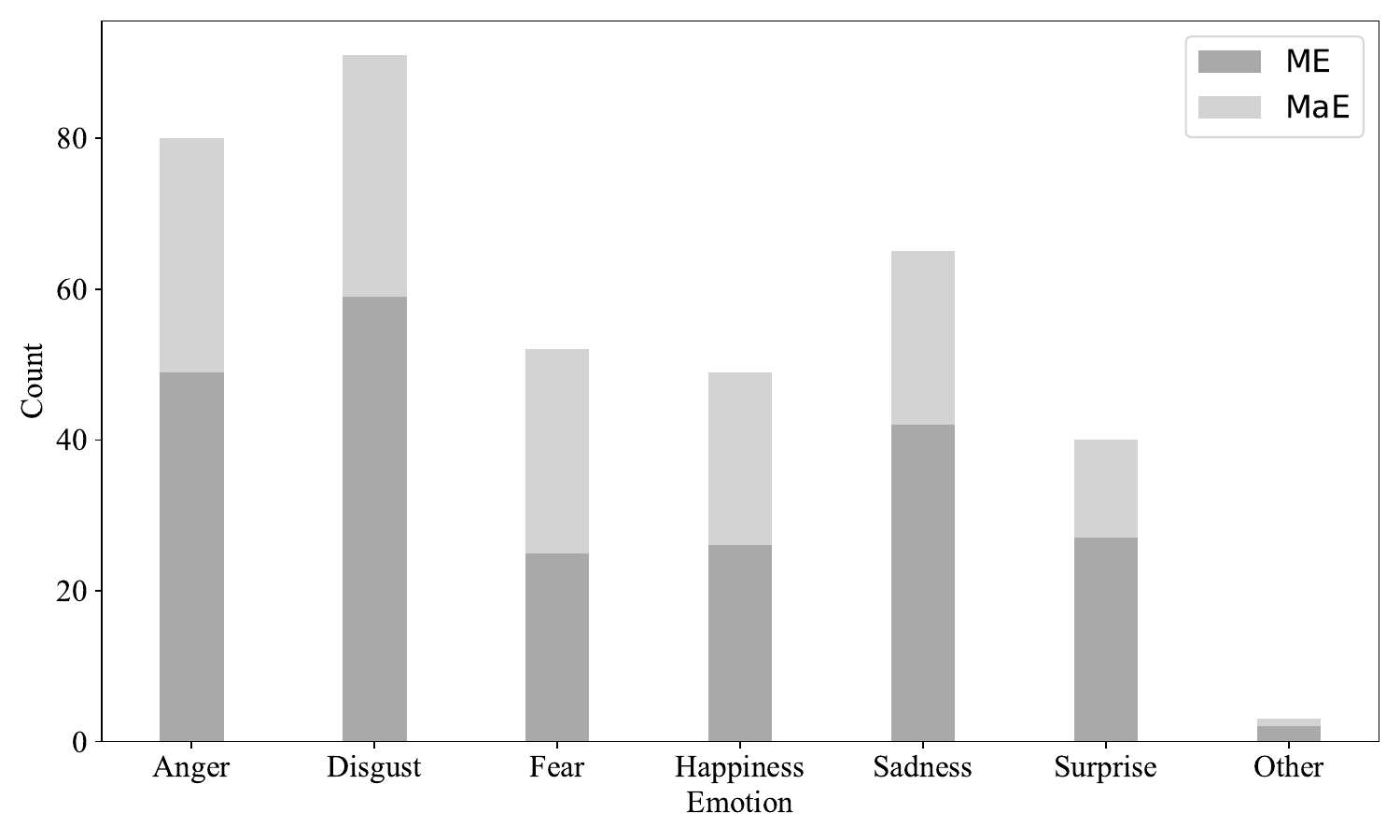}
    \caption{\highlight{Graphical presentation of acquired data distribution with respect to each emotion categories}}
    \label{fig:emoHisto}
\end{figure}

\highlight{Secondly, for emotional annotation, the coders need to make judgments by integrating the AU labels and the emotional type of the eliciting material. The two coders separately annotated the emotions of 380 expressions with coded AUs. Under a seven-category standard of six basic emotions and others (emotions ambiguous or difficult to classify into the six basic categories), the consistency of emotional coding was as high as 99\%. The final consensus coding results, along with the coding statistics, are shown in Table.~\ref{tab:emotion_data} and Fig.~\ref{fig:emoHisto}.}
\par
\highlight{During the data collection process, participants were instructed to maintain a neutral expression while watching emotionally stimulating materials, resulting in a slightly higher number of MEs compared to MaEs in CASMEMG. Additionally, we used 24 elicitation videos, more than other datasets and almost twice as many as CAS(ME)$^3$ (13 elicitation videos), to balance the six basic emotions as much as possible. This ensures a relatively even distribution of emotional samples in our database. Furthermore, Table.~\ref{tab:emotion_data} shows the correlation between AUs and emotions, such as AU4 and AU7 with negative emotions, and AU12 with happiness. Additionally, anger and disgust can be differentiated by (AU16, AU23) and (AU9, AU17) respectively. Fear, another challenging negative emotion to distinguish, can be identified by focusing on AU5. Sadness and surprise each have their own distinctive AUs. Utilizing these patterns can further enhance the performance of intelligent ME analysis based on expert knowledge.}

\begin{table*}[]
    \centering
\begin{tabular}{|c|c|c|c|c|>{\highlight}c|}
\hline
        \highlight{\textbf{Emotion}} & \highlight{\textbf{ME}} & \highlight{\textbf{MaE}} & \highlight{\textbf{Total \#}} & \highlight{\textbf{Dominant AUs}} & \highlight{\textbf{Main EMG channels}} \\
        \hline
        \highlight{Anger} & \highlight{49} & \highlight{31} & \highlight{80} & \highlight{AU4, AU7, AU16, AU23} & \highlight{c2, c3, c6} \\
        \hline
        \highlight{Disgust} & \highlight{59} & \highlight{32} & \highlight{91} & \highlight{AU4, AU7, AU9, AU17} & \highlight{c2, c3, c4, c6} \\
        \hline
        \highlight{Fear} & \highlight{25} & \highlight{27} & \highlight{52} & \highlight{AU4, AU5, AU7} & \highlight{c2, c3} \\
        \hline
        \highlight{Happiness} & \highlight{26} & \highlight{23} & \highlight{49} & \highlight{AU6, AU12, AU14} & \highlight{c3, c5, c7} \\
        \hline
        \highlight{Sadness} & \highlight{42} & \highlight{23} & \highlight{65} & \highlight{AU14, AU15, AU24, AU43} & \highlight{c3, c6, c7} \\
        \hline
        \highlight{Surprise} & \highlight{27} & \highlight{13} & \highlight{40} & \highlight{AU2, AU5} & \highlight{c1, c3} \\
        \hline
        \highlight{Other} & \highlight{2} & \highlight{1} & \highlight{3} & \highlight{AU45} & \highlight{c3} \\
        \hline
\end{tabular}
\caption{\highlight{The statistical classification of seven emotional categories}}
\label{tab:emotion_data}
\end{table*}
\subsection{\highlight{Baseline method on EMG-based ME interval spotting}}

\highlight{We proposed a facial expression interval detection method based on EMG peak and trough detection, and compared it with manually annotated results to demonstrate the potential of semi-automatic coding based on EMG. (Code is available on \url{https://github.com/MELABIPCAS/CASMEMG.git}).}
\par
\highlight{Regarding the methodology, the EMG signal from the channel with the largest amplitude among all channels was selected and normalized. The ratio of the mean value of the normalized EMG data $E_n$ is used to define the threshold scale, $T_r$. This threshold scale is given by:}

\begin{equation}
\highlight{T_r = k \times \text{mean}(E_{n})}
\end{equation}
\highlight{where $k$ is a scaling factor, and $E_{n}$ represents the normalized EMG data.}
\par
\highlight{A sliding window approach is used to evaluate the EMG data $E_{n}$. The number of sampling points in each window is denoted as $W_L$. The step size, or the number of sampling points by which the window moves forward or backward, is denoted as $S_L$. For each window position, the EMG energy is calculated. Windows that exceed the threshold scale $T_r$ are marked with an indicator. This indicates that the average energy within the window is above the threshold. If the number of consecutive indicators exceeds the threshold $S_n$, it suggests a peak in the EMG data. This implies a potential facial movement within this segment composed of these consecutive windows. This can be mathematically expressed as:}
\begin{equation}
\highlight{\sum_{i=n}^{n+S_L-1} \mathbf{1}(x_i > T_r) \geq S_n}
\end{equation}
\highlight{where $\mathbf{1}(\cdot)$ is the indicator function that equals 1 if the condition inside is true and 0 otherwise.}
\par
\highlight{Once an EMG peak indicating the presence of a facial movement is identified, the onset and offset times of the facial movement are determined. The algorithm defines the troughs before and after the EMG peak as the start and end times of the movement, respectively. }

\highlight{For the \textbf{Onset Time}:}

\begin{equation}
\highlight{\text{Onset Time} = \arg\min_{t \in [p-W_f, p]} EMG(t)}
\end{equation}

\highlight{where $W_f$ represents the number of windows moved backward, and $p$ is the position of the peak.}
\par
\highlight{In contrast, for the \textbf{Offset Time}:}
\begin{equation}
\highlight{\text{Offset Time} = \arg\min_{t \in [p, p+W_b]} EMG(t)}
\end{equation}

\highlight{where \(W_b\) represents the number of windows moved forward. The identified minimum values (troughs) within these ranges are recorded to provide the onset and offset times of the facial expressions.}

\subsection{\highlight{Experimental result and analysis}}
\highlight{During the algorithm validation process, the EMG segments were derived from annotated MEs. Specifically, we extended the annotated expression times by 0.5 seconds before and after within the complete EMG signal. The validation of the EMG detection algorithm involves pinpointing the start and end times of MEs within these segments. This serves two purposes: demonstrating that EMG can effectively help coders accurately locate key time points of MEs within roughly identified regions, and providing a baseline result for an EMG-based expression interval detection algorithm in the EMG database.}
\par
\highlight{A total of 380 EMG facial expression samples were used in the experiment. In cases where multiple values appear in the onset and offset detection results, the final onset and offset values used to determine the key moments of MEs are the maximum and minimum values, respectively. }
\par
\highlight{We conducted 900 iterations to explore the parameters involved in the algorithm. This process identified the optimal parameter combination: $W_L$ = 60, $S_L$ = 30, $k$ = 1, $S_n$ = 5, $W_f$ = 2, $W_b$ = 6.}
\par
\highlight{As listed in Table~\ref{tab:evaluation_metrics}, we analyzed the results at both the interval and moment levels. At the interval level, there were only 8 segments where the EMG detection algorithm failed to detect the corresponding interval. Additionally, we compared the overlap (IoU) rates for all detected intervals. We found that 95.3\% of the intervals had an overlap rate greater than 0.5 (among 380 EMG segments). The average overlap rate for all intervals was 84.7\%.}
\par
\highlight{We then evaluated the performance of onset and offset position estimation. We calculated the differences between the predicted onset and offset positions and the ground truth. The table lists the mean, Standard Error (SE) and Root Mean Square Error (RSME) values. The results show that EMG can provide accurate predictions for the start and end times of expressions. It is also expected that the SE and RSME values of the onset moment are better than those of the offset moment. This is because the start of an action is usually regular, typically transitioning from weak to strong. However, at the end of an action, it does not always return to a neutral state, making accurate detection challenging as it may not form a clear trough. This phenomenon is consistent with the findings from the previous research~\cite{yan2015quantifying}.}

\begin{table}[h!]
    \centering
    \caption{\highlight{EMG-based facial expression interval detection performance. N\_nan represents the number of EMG segments in which expression intervals were not detected. Mean\_IoU and P(IoU)$>$0.5 indicate the average overlap rate and the proportion of overlap rates greater than 0.5. }}
    \label{tab:evaluation_metrics}
       \begin{tabular}{|c|c|c|c|}
        \hline
        \multicolumn{2}{|c|}{\multirow{3}{*}{\highlight{\textbf{Interval level}}}} & \highlight{\textbf{N\_nan}} & \highlight{8} \\
        \cline{3-4}
        \multicolumn{2}{|c|}{} & \highlight{\textbf{Mean\_IoU}} & \highlight{0.847} \\
        \cline{3-4}
        \multicolumn{2}{|c|}{} & \highlight{\textbf{P(IoU $>$ 0.5)}} & \highlight{0.953} \\
        \hline
        \multirow{6}{*}{\highlight{\textbf{Moment-level}}} & \multirow{3}{*}{\highlight{$\Delta$Onset}} & \highlight{\textbf{Mean}} & \highlight{0.206s} \\
        \cline{3-4}
        & & \highlight{\textbf{SE}} & \highlight{0.005s} \\
        \cline{3-4}
        & & \highlight{\textbf{RMSE}} & \highlight{0.230s} \\
        \cline{2-4}
        & \multirow{3}{*}{\highlight{$\Delta$Offset}} & \highlight{\textbf{Mean}} & \highlight{0.203s} \\
        \cline{3-4}
        & & \highlight{\textbf{SE}} & \highlight{0.012s} \\
        \cline{3-4}
        & & \highlight{\textbf{RMSE}} & \highlight{0.309s} \\
        \hline
    \end{tabular}
\end{table}

\section{Conclusion and Perspective}

To address the lack of empirical studies on the physiological characteristics, we characterized MEs via EMG signals. The study showed that MEs could be characterized as facial expressions that are short in duration, low in intensity, and spontaneous (unaware and uncontrollable). 
First, this study accurately estimated the ME duration interval, with the main duration ranging from 307 to 327 ms. This duration range is a more precise time interval estimation of ME duration compared with the duration of less than 1/2~s. Second, we demonstrated that the intensity of MEs was significantly smaller than that of MaEs and then provided an estimate of the intensity interval of MEs. The maximum EMG signal amplitude of MEs ranged from 7.01$\%$ to 9.21$\%$ of the MVC. In other words, the amplitude of muscle movements during ME production is measured through a specific quantitative indicator, i.e., MVC\%, which is a preliminary attempt to investigate the physiological mechanisms behind ME production in the future. 
Third, we also explored the internal characteristic, i.e., the cognitive dimension of the ME expression process. The results of the keystroke response and self-report studies suggest that MEs are unaware and uncontrollable. 

Overall, this is the first in-depth and objective exploration of ME external characteristics by EMG. Along with a further internal characteristic analysis, this study lays the foundation for subsequent studies of ME physiological mechanisms. Furthermore, this work have guiding importance for the algorithm design of automated ME analysis. These characteristics can be used as the valuable priori knowledge for feature learning, thus improving the network's ability to analyze MEs. \highlight{Additionally, we have released CASMEMG, the first ME database that includes EMG signals, facilitating deeper exploration of micro-expression movement mechanisms.}
\par
In addition to providing an overview of our work, this section also aims to discuss the limitations encountered in our research and explore some promising techniques. These include the ME characteristics such as asymmetry, the ecological validity of the ME elicitation paradigm, and semi-automatic EMG-based ME coding.

\par

\textbf{Asymmetrical MEs:} As mentioned at the beginning of Section~\ref{sec:result}, twenty-one expression samples were excluded due to the absence of corresponding EMG data. If the ME occurs only on the right side of the face, we can code the expression but cannot capture the corresponding EMG signal. Although we aimed to balance facial EMG signal acquisition and expression coding, our method still has room for improvement. Future work could explore the design of flexible transparent electrodes~\cite{inzelberg2018printed} tailored for facial EMG acquisition and optimized electrode placement along the facial contour.

\par
\textbf{High ecological validity elicitation paradigm:} Due to the ``expressionless" elicitation paradigm, facial movements with large intensities in our experiment were suppressed, the intensity of our collected MaEs reached only 23$\%$ of MVC. There may be differences in intensity between passively evoked expressions in the laboratory and actively flowing expressions in interpersonal interactions~\cite{hess1995intensity,porter2012secrets}. In future studies, we will collect facial expressions in interaction situations for comparative analysis with current data.

\par



\textbf{Semi-automatic EMG-based ME coding:} Capturing all muscle actions in an ME with the naked eye is challenging, and traditional manual coding is prone to self-report and coder biases. In contrast, facial EMG can detect subtle muscle activations more accurately and objectively, reducing these biases \highlight{(see the user study in Appendix~\ref{app:userstudy})}. However, EMG signals also capture non-emotional movements, making pre-processing difficult. \highlight{Combining automatic EMG-based interval detection with manual annotation} could enhances the accuracy of ME coding. This kind of semi-automatic coding could accelerates the creation of large ME databases, facilitating improved intelligent analysis and exploration of physiological mechanisms.

\ifCLASSOPTIONcaptionsoff
  \newpage
\fi

\appendices
\section{\highlight{Correspondence between EMG channels, AUs, and emotions} }
\label{app:channel}
\highlight{Table~\ref{tab:au_table} lists the relationships between facial muscle channels and AUs, while Table~\ref{tab:emotion_au_table} examines the primary muscle movements corresponding to different emotions.}
\begin{table}[h]
\centering
\caption{\highlight{Correspondence between EMG channels and AUs}}
\begin{tabular}{|c|c|c|}
\hline
\highlight{\textbf{ID}} & \highlight{\textbf{Muscle} }& \highlight{\textbf{AUs} }\\ \hline
        \highlight{c1} & \highlight{Frontalis} & \highlight{AU1, AU2}  \\ \hline
        \multirow{2}{*}{\highlight{c2}} & \highlight{Corrugator supercilii} & \multirow{2}{*}{\highlight{AU4}}  \\ 
        & \highlight{supercilii} &  \\\hline
        \multirow{2}{*}{\highlight{c3}} & \multirow{2}{*}{\highlight{Orbicularis oris}} & \highlight{AU5, AU6, AU7, AU4, AU41,}   \\ 
        &  & \highlight{AU42, AU43, AU44, AU45}  \\\hline
        \multirow{2}{*}{\highlight{c4}} & \highlight{Levator labii}  & \multirow{2}{*}{\highlight{AU9, AU10}}  \\ 
        & \highlight{superioris alaeque nasi} & \\\hline
        \highlight{c5} & \highlight{Zygomaticus major} & \highlight{AU11, AU12}  \\ \hline
        \multirow{2}{*}{\highlight{c6}} & \multirow{2}{*}{\highlight{Orbicularis oris}} & \highlight{AU15, AU16, AU17, AU18,} \\
        & & \highlight{AU22, AU23, AU24, AU28}  \\ \hline
        \highlight{c7} & \highlight{Depressor anguli oris} & \highlight{AU14, AU25}  \\ \hline
\end{tabular}

\label{tab:au_table}
\end{table}

\begin{table*}[h]
\centering
\caption{\highlight{Relationship between emotions, AUs and EMG channels}}
\begin{tabular}{|c|c|c|c|}

        \hline
        \highlight{\textbf{Emotion}} & \highlight{\textbf{Action Units (AUs)}} & \highlight{\textbf{Channels}} & \highlight{\textbf{Manifestation}} \\ \hline
        \highlight{Happy} & \highlight{AU6, AU12} & \highlight{c3, c5} & \highlight{Smile: Pulls the corner of the mouth back or upward.} \\ \hline
        \highlight{Sad} & \highlight{AU45, AU15} & \highlight{c3, c7} & \highlight{Grinning; Eyes closing.} \\ \hline
        \highlight{Fear} & \highlight{AU4, AU7} & \highlight{c2, c3} & \highlight{Frowning and eye contraction.} \\ \hline
        \highlight{Surprise} & \highlight{AU2, AU5} & \highlight{c1, c3} & \highlight{Widening eyes and raising eyebrows.} \\ \hline
        \highlight{Disgust} & \highlight{AU4, AU9, AU17} & \highlight{c2, c4, c6} & \highlight{Frowning, nose twist, and curling lips.} \\ \hline
        \highlight{Angry} & \highlight{AU4, AU7} & \highlight{c2, c3} & \highlight{Frowning and eye widening.} \\ \hline
\end{tabular}

\label{tab:emotion_au_table}
\end{table*}

\section{\highlight{User Study on EMG-based Coding Process}}
\label{app:userstudy}
\highlight{We conducted a user study comparing purely manual coding with and without the aid of EMG. This means that no algorithms were involved in detecting changes in EMG or visual signals; the process relied entirely on the observations of the coders. Specifically, the study involves two participants: one experienced FACS-certified coder and one novice coder who received brief training (presented with 10 ME samples). Regarding the video to be coded, we selected 12 videos with an average duration of 124 seconds. These videos had previously been coded for the occurrence times (onset and offset) of MaEs/MEs by two experienced coders based on EMG and video data, serving as the ground truth for this experiment. Ten of these videos contained only one expression each, while the other two videos each contained four expressions. Detailed information about these videos is provided in Table~\ref{table:comparison}.}
\par
\begin{figure}
    \centering
    \includegraphics[width=\linewidth]{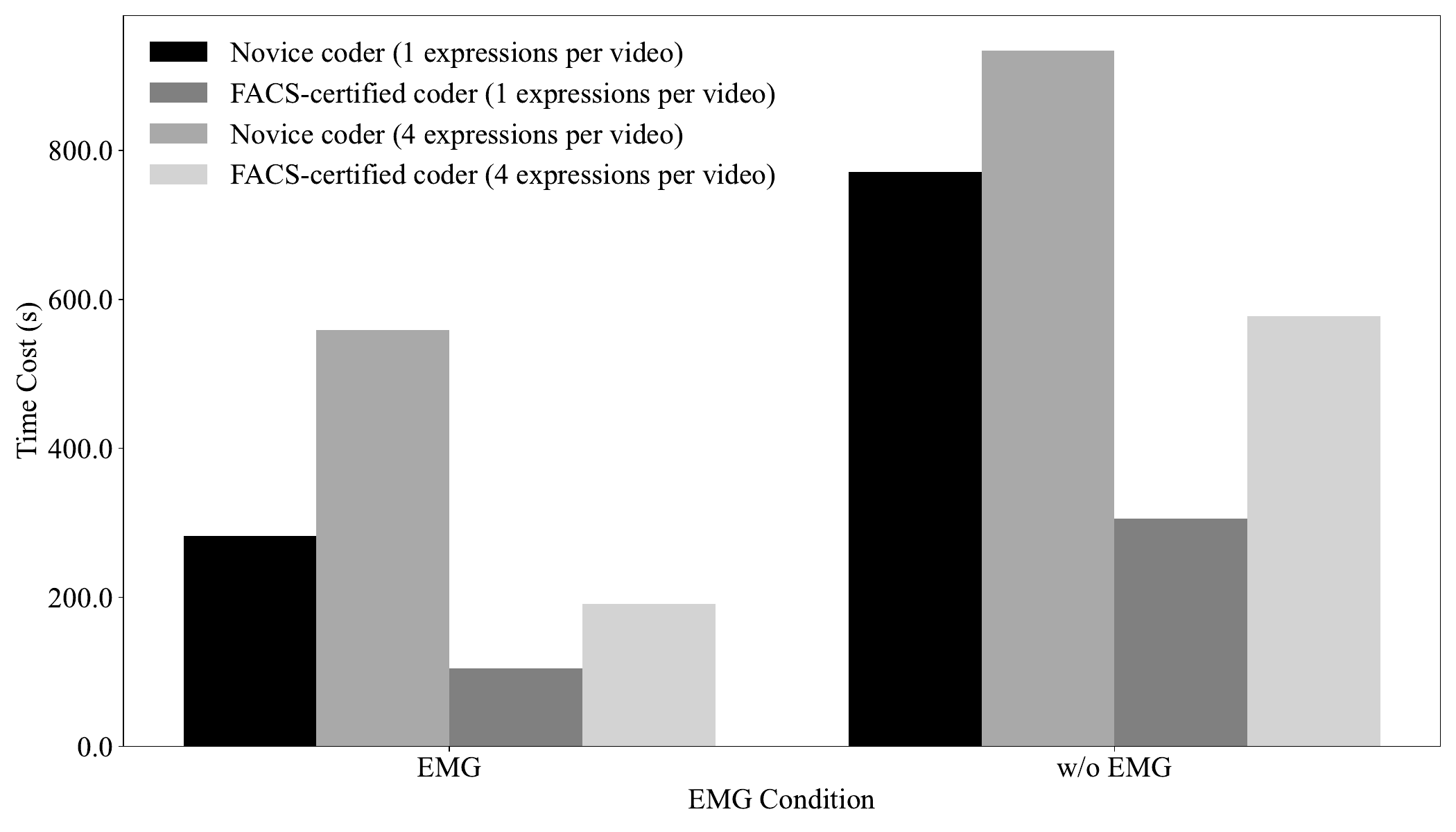}
    \caption{\highlight{Average encoding times under various conditions}}
    \label{fig:codingavg}
\end{figure}

\highlight{During the preliminary experiment, we discovered that novice coders found it challenging to complete the coding task within 20 minutes without any hints. Therefore, without compromising the experimental goal (i.e., proving the effectiveness of EMG in assisting ME coding), we reduced the difficulty of the coding task. The number of expressions in each video was explicitly informed to the participants, and they were only required to accurately code the onset and offset times of the expressions they identified within 20 minutes.
Table~\ref{table:comparison} presents the time cost comparison of the two participants’ encoding performance with and without the aid of EMG. Fig.~\ref{fig:codingavg} shows the average encoding times for different types of coders under various conditions. It is evident that with the aid of EMG, both experienced and novice coders experienced a significant increase in coding efficiency. Specifically, for videos containing a single expression, the improvements for experienced and novice coders were 65.86\% and 63.35\%, respectively. For videos with multiple expressions, the improvements were 66.89\% for experienced coders and 40.15\% for novice coders. Besides, in post-study interviews, both experienced and novice coders reported that the EMG system greatly assisted their coding efforts. It not only enhanced their coding efficiency but also boosted their confidence in coding accuracy, making the system user-friendly.}

\begin{table}[ht]
\centering
\caption{\highlight{Coding efficiency comparison with and without EMG. $n_e$ denotes the number of facial expressions in one video to be coded. $t_{N}$ and $t_{F}$ represent time costs for novice and FACS-certified coders, respectively. }}
\begin{tabular}{|c|c|c|c|c|c|}
\hline
        \highlight{EMG} & \highlight{\textbf{$n_{e}$}} & \highlight{\textbf{video index}} & \highlight{\textbf{duration (s)}} & \highlight{\textbf{$t_{N}$ (s)}} & \highlight{\textbf{$t_{F}$ (s)}} \\ \hline
        \multirow{6}{*}{\highlight{with}} & \multirow{5}{*}{\highlight{1}} & \highlight{7-4} & \highlight{83} & \highlight{377.023} & \highlight{104.029} \\ \cline{3-6}
        & & \highlight{8-2} & \highlight{78} & \highlight{345.012} & \highlight{72.009} \\ \cline{3-6}
        & & \highlight{10-16} & \highlight{229} & \highlight{298.028} & \highlight{176.019} \\ \cline{3-6}
        & & \highlight{11-1} & \highlight{153} & \highlight{193.017} & \highlight{94.005} \\ \cline{3-6}
        & & \highlight{13-1} & \highlight{153} & \highlight{200.023} & \highlight{76.015} \\ \cline{2-6}
        & \highlight{4} & \highlight{5-11} & \highlight{149} & \highlight{559.016} & \highlight{191.048} \\ \hline
        \multirow{6}{*}{\highlight{w/o}} & \multirow{5}{*}{\highlight{1}} & \highlight{7-6} & \highlight{78} & \highlight{735.031} & \highlight{225.017} \\ \cline{3-6}
        & & \highlight{8-3} & \highlight{52} & \highlight{846.097} & \highlight{182.025} \\ \cline{3-6}
        & & \highlight{10-13} & \highlight{85} & \highlight{486.035} & \highlight{252.056} \\ \cline{3-6}
        & & \highlight{11-12} & \highlight{167} & \highlight{904.071} & \highlight{469.031} \\ \cline{3-6}
        & & \highlight{13-17} & \highlight{104} & \highlight{884.075} & \highlight{381.003} \\ \cline{2-6}
        & \highlight{4} & \highlight{5-5} & \highlight{154} & \highlight{934.02} & \highlight{577.012} \\ \hline
\end{tabular}
\label{table:comparison}
\end{table}

\par
\highlight{This improvement can be attributed to the following reasons: Without the aid of EMG, coders needed to meticulously watch the entire video, often in slow motion. However, with EMG, coders could quickly pinpoint the locations of expressions based on changes in EMG signals. Additionally, it is noteworthy that the improvement was more pronounced for experienced coders. While EMG provides potential time frames for facial movements, a certain level of experience is required to filter these candidate segments and accurately time-code the facial movements. This highlights the importance of familiarity with facial action coding.}
\par
\highlight{It is important to note that conducting large-scale user studies for ME coding is very challenging. On one hand, there are very few experienced coders with FACS certification, which requires extensive knowledge of facial muscle movement patterns and often a background in psychology for accurate coding. On the other hand, ME coding, especially without EMG assistance, is extremely time-consuming. For novice coders, ME coding is particularly difficult. In our table, we only recorded the time taken by coders to encode a single video. In reality, the novice coder took a total of 4 hours to complete the coding. This was partly due to the time required to thoroughly study the ME sample examples. Additionally, the novice coder found that coding MEs in a video demanded intense concentration on the screen, necessitating breaks after coding each video before proceeding to the next. These factors limited our ability to conduct large-scale user studies. However, we believe that the coding experiments conducted with these two participants are representative and demonstrate the effectiveness and user-friendliness of the EMG-assisted coding system.}



\bibliographystyle{IEEEtran}
\bibliography{reference}
%

%

\begin{IEEEbiography}[{\includegraphics[width=1in,height=1.25in,clip,keepaspectratio]{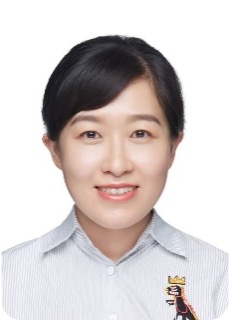}}]{Jingting LI}
is currently an associate researcher at the Institute of Psychology, Chinese Academy of Sciences (CAS). She received the PhD degree in Signal, Image, Vision from CentraleSupélec in 2019. She served as the chair of the ACMMM'21, 22 and 23 FME workshop and MEGC Grand challenge, organized and hosted several China Society of Image and Graphics (CSIG) online 
ME workshop sessions. Her current research interests include image processing, computer vision and pattern recognition.
\end{IEEEbiography}

\begin{IEEEbiography}[{\includegraphics[width=1in,height=1.25in,clip,keepaspectratio]{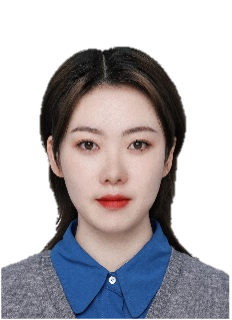}}]{Shaoyuan Lu}
received the B.S. degree in Psychology major from Anhui Normal University, China, in 2020. She is currently pursuing the M.Psy. degree in the Institute of Psychology, Chinese Academy of Sciences. Her current research interests include facial ME, visual psychophysics and neurophysiology.
\end{IEEEbiography}

\begin{IEEEbiography}[{\includegraphics[width=1in,height=1.25in,clip,keepaspectratio]{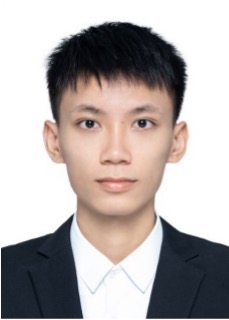}}]{Yan Wang}
received the B.S. degree in Psychology major from Anhui Normal University, China, in 2020. He is currently pursuing the M.Psy. degree in the Institute of Psychology, Chinese Academy of Sciences. Her current research interests include facial ME, visual psychophysics and neurophysiology.
\end{IEEEbiography}


\begin{IEEEbiography}[{\includegraphics[width=1in,height=1.25in,clip,keepaspectratio]{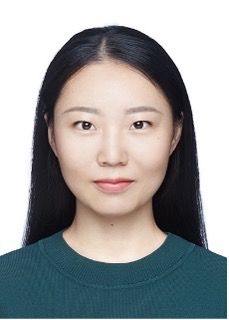}}]{Zizhao Dong}
received the B.S.Ed degree in applied psychology major from China Women's University, China, in 2018. She is currently pursuing the M.Psy. degree in the Institute of Psychology, Chinese Academy of Sciences. Her current research interests include facial ME , visual psychophysics and neurophysiology.

\end{IEEEbiography}

\begin{IEEEbiography}[{\includegraphics[width=1in,height=1.25in,clip,keepaspectratio]{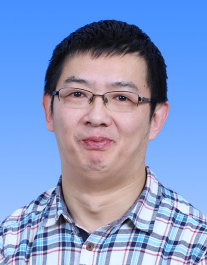}}]{Su-Jing Wang}
(M'12-SM'19) is an Associate Researcher, PhD supervisor at the Institute of Psychology, CAS. He received the Ph.D degree from the College of Computer Science and Technology of Jilin University in 2012. His current research interests include pattern recognition and machine learning.
He won the first prize of the 8th Wu Wenjun Artificial Intelligence Science and Technology Award in 2018. He was selected as one of the top 2\% of scientists in the world in 2020 for "Impact of the Year". 
\end{IEEEbiography}

\begin{IEEEbiography}[{\includegraphics[width=1in,height=1.25in,clip,keepaspectratio]{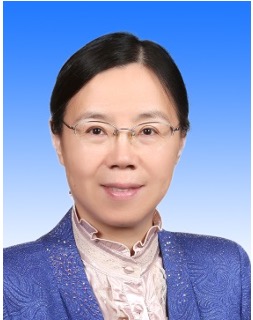}}]{Xiaolan Fu}
(M’13) received her Ph. D. degree in 1990 from Institute of Psychology, Chinese Academy of Sciences. Currently, she is a Senior
Researcher at Cognitive Psychology. Her research interests include visual and computational cognition: (1) attention and perception, (2) learning and memory, and (3) affective computing. At present, she is the director of Institute of Psychology, Chinese Academy of Sciences and the director of department of psychology, University of the Chinese Academy of Sciences.
\end{IEEEbiography}




\end{document}